\documentclass[longauth]{aa}
\usepackage{booktabs}
\usepackage{siunitx}
\usepackage{graphicx,txfonts}
\usepackage[colorlinks=true, allcolors=blue]{hyperref}
\usepackage{natbib}

\bibpunct{(}{)}{;}{a}{}{,}

\newcommand{\gt}{>}
\newcommand\identity{1\kern-0.25em\text{l}}
\newcommand{\Grp}{\ensuremath{G_\mathrm{RP}}}
\newcommand{\Kmag}{\ensuremath{\mathrm{m}_K}}
\newcommand{\Vmag}{\ensuremath{\mathrm{m}_V}}
\newcommand{\Rmag}{\ensuremath{\mathrm{m}_R}}
\newcommand{\Msun}{\ensuremath{\mathrm{M}_\odot}}
\newcommand{\component}[1]{{#1}}

\title{First light for the GRAVITY+ Adaptive Optics:\\extreme adaptive optics for the Very Large Telescope Interferometer}
\titlerunning{First light for GRAVITY+ Adaptive Optics}
\authorrunning{GRAVITY+ Collaboration}

\author{GRAVITY+ Collaboration\thanks{GRAVITY+ is developed in a collaboration by the Max Planck Institute for extraterrestrial Physics, the Institute National des Sciences de l'Univers du CNRS (INSU) with its institutes LIRA~/~Paris Observatory-PSL, IPAG~/~Grenoble Observatory, Lagrange~/~Cote d'Azur Observatory and CRAL~/~Lyon Observatory, the Max Planck Institute for Astronomy, the University of Cologne, the CENTRA - Centro de Astrofisica e Gravita\c c\~ao, the University of Southampton, the Katholieke Universiteit Leuven, University College Dublin, Universidad Nacional Aut\'onoma de M\'exico and the European Southern Observatory. \newline
Corresponding authors: J.-B. Le Bouquin (email: \href{mailto:jean-baptiste.lebouquin@univ-grenoble-alpes.fr}{jean-baptiste.lebouquin@univ-grenoble-alpes.fr}), G. Bourdarot (email: \href{mailto:bourdarot@mpe.mpg.de}{bourdarot@mpe.mpg.de}), A. Berdeu (email: \href{mailto:anthony.berdeu@obspm.fr}{anthony.berdeu@obspm.fr})
}
R.~Abuter\inst{8}
\and F.~Allouche\inst{23, 9}
\and C.~Bailet\inst{23}
\and M.~Benisty\inst{3}
\and A.~Berdeu\inst{2,9}
\and J.-P.~Berger\inst{5}
\and P.~Berio\inst{23}
% ...
}

\institute{
Max Planck Institute for extraterrestrial Physics,
Giessenbachstra{\ss}e~1, 85748 Garching, Germany
\and LIRA, Observatoire de Paris, Universit\'e PSL, Sorbonne Universit\'e, Universit\'e Paris Cit\'e, CY Cergy Paris Universit\'e, CNRS, 92190 Meudon, France
\and Max Planck Institute for Astronomy, K\"onigstuhl 17, 
69117 Heidelberg, Germany
\and $1^{\rm st}$ Institute of Physics, University of Cologne,
Z\"ulpicher Stra{\ss}e 77, 50937 Cologne, Germany
\and Univ. Grenoble Alpes, CNRS, IPAG, 38000 Grenoble, France
\and Universidade de Lisboa - Faculdade de Ci\^encias, Campo Grande,
1749-016 Lisboa, Portugal 
\and Faculdade de Engenharia, Universidade do Porto, rua Dr. Roberto
Frias, 4200-465 Porto, Portugal 
\and European Southern Observatory, Karl-Schwarzschild-Stra{\ss}e 2, 85748
Garching, Germany
\and European Southern Observatory, Casilla 19001, Santiago 19, Chile
\and Sterrewacht Leiden, Leiden University, Postbus 9513, 2300 RA
Leiden, The Netherlands
\and Departments of Physics and Astronomy, Le Conte Hall, University
of California, Berkeley, CA 94720, USA
\and CENTRA - Centro de Astrof\'{\i}sica e
Gravita\c c\~ao, IST, Universidade de Lisboa, 1049-001 Lisboa,
Portugal
\and Department of Physics, Technical University Munich, James-Franck-Stra{\ss}e 1,  85748 Garching, Germany
\and Max Planck Institute for Radio Astronomy, Auf dem H\"ugel 69, 53121 Bonn, Germany
\and Department of Physics, University of Illinois, 1110 West Green Street, Urbana, IL 61801, USA
\and Hamburger Sternwarte, UniversitÃ¤t Hamburg, Gojenbergsweg 112, 21029 Hamburg, Germany
\and School of Physics \& Astronomy, University of Southampton, Southampton, SO17 1BJ, United Kingdom
\and Center for Astrophysics | Harvard \& Smithsonian, 60 Garden Street, Cambridge, MA, 02138, USA
\and Max Planck Institute for Astrophysics, Karl-Schwarzschild-Stra{\ss}e 1, 85741 Garching, Germany
\and Institute of Astronomy, KU Leuven, Celestijnenlaan 200D, B-3001, Leuven, Belgium
\and School of Physics, University College Dublin, Dublin 4, Belfield, Ireland
\and Instituto de Astronomia, Universidad Nacional Aut\'onoma de M\'exico, Apdo. Postal 70264, Ciudad de Mexico 04510, Mexico
\and Laboratoire Lagrange, Univ. CÃ´te d'Azur, CNRS, OCA, Nice, France 
% ...
}

\date{Received: 26 May, 2025, 2025; accepted: 24 September, 2025}

\abstract{
GRAVITY+ improves by orders of magnitude the sensitivity, sky-coverage, and contrast of the Very Large Telescope Interferometer (VLTI). A central part of this project is the development of Gravity Plus Adaptive Optics (GPAO), a dedicated high-order and laser-guide star adaptive optics (AO) system for VLTI. GPAO consists of four state-of-the-art AO systems that equip all 8m class Unit Telescopes (UTs) for the wavefront correction of the VLTI instruments. It offers both visible and infrared natural guide star (NGS) and laser guide star (LGS) operations. The paper presents the design, operations, and performances of GPAO. We illustrate the improvement brought by GPAO with interferometric observations obtained during the commissioning of the NGS mode at the end of 2024. These science results include the first optical interferometry observations of a redshift $z\sim4$ quasar, the spectroscopy of a cool brown-dwarf with magnitude $K\sim 21.0$, the first observations of a Class I young star with GRAVITY, and the first sub-micro arcsecond differential astrometry in the optical. Together with the entire GRAVITY+ project, the implementation of GPAO is a true paradigm shift for observing the optical Universe at very high angular resolution.
}


\begin{document} 

\maketitle
\section{Introduction}
Adaptive optics (AO) was identified early on as a prerequisite for interferometry on large telescopes, in order to work with apertures larger than the Fried diameter and increase the coherent volume for interferometry \citep{1988ESOC...29..899L, Eisenhauer2023}. The use of AO directly results in a sensitivity improvement of $\propto D^4$ for diffraction-limited interferometric observations, compared to $\propto D^{1/3}$ in the case of seeing-limited observations \citep{Eisenhauer2023}.
As such, the use AO has been rapidly proposed to correct the wavefront before interferometric combination in place of multi-speckle interferometry \citep{Roddier1981}, as well as to correct the coupling of light into single-mode fibers \citep{Shaklan1988} for infrared interferometry.
\newline

In this paper, we present the Gravity Plus Adaptive Optics (GPAO) system, the new AO system of the Very Large Telescope Interferometer (VLTI), and its first on-sky observations. Section \ref{sec:ao_at_vlti} introduces the instrumental context at the VLTI and the rationale for a new system. Section \ref{sec:design} details the design choices and the implementation of GPAO in the VLTI \SI{8}{\meter} telescopes. Section \ref{sec:op_and_perf} presents the operation and the performance of AO. Section \ref{sec:science} illustrates the new science enabled by presenting the first interferometric observations obtained with the natural guide star mode of GPAO that were commissioned at the end of 2024. The laser-assisted mode will be commissioned at the end of 2025 and its on-sky results will be presented in a later paper. All acronyms used throughout this paper are described in Appendix~\ref{app:nomenclature}.

\section{Context at the VLTI and GRAVITY+}
\label{sec:ao_at_vlti}

\subsection{Optical-Infrared interferometry with large telescopes and adaptive optics:}

The role of AO for the 8m-class telescopes on the Very Large Telescope Interferometer (VLTI) was emphasized from its initial design: "The largest gains in sensitivity of the VLTI will come from the incorporation of AO which will make the telescope diffraction limited at \SI{2}{\micro\meter}" \citep{Beckers1990}.
Therefore, a dedicated AO facility was implemented at VLTI from its early start with the Multi-Application Curvature Adaptive Optics \citep[MACAO,][]{Arsenault2003}. The emphasis was made on multipurpose applications, with a trade-off between limiting magnitude and wavefront quality in the absence of laser guide stars and technological limitations at the time. The VLTI was later upgraded with an infrared AO, the GRAVITY Coud\'e Infrared Adaptive Optics \citep[CIAO,][]{Scheithauer2016}, to provide AO for the extremely red Galactic Center region.

The next step on large telescopes is the development of an AO facility enabling both natural guide star (NGS) and laser guide star (LGS) operations. Unlike small telescopes, for which the limiting magnitude is set by the number of photons available for fringe-tracking, the limiting magnitude for large telescopes is constrained by the AO system~\citep{Eisenhauer2023}. This originates from the fact that the number of photons available for fringe-tracking increases with the collecting area, while for AO this number is set by the number of photons per turbulence cell, whatever the size of the telescope. For a telescope size larger than the Fried parameter, the limiting magnitude is thus independent of the size of the telescope.  This limitation is lifted with LGS AO, where the high-order correction is done on an artificial star. The limiting magnitude in this case is set by the tip-tilt correction, which requires a nearby star in the isokinetic patch. This magnitude is comparable to the fringe-tracking limiting magnitude. Therefore, an LGS AO system is the only way to reach the ultimate interferometric sensitivity of an array of large apertures. Moreover, when LGS operations are available to cover the faint end, the NGS operations can be optimized for peak performances on the bright end, overcoming the trade-off between limiting magnitude and wavefront quality.

\subsection{GRAVITY overview}

The GRAVITY instrument \citep{2017A&A...602A..94G} is a four beam interferometric instrument, offering milli-arcsecond (mas) resolution imaging for objects as faint as $\Kmag>20$, together with $30-100$ micro-arcsecond (\SI{}{\micro as}) narrow-angle astrometry, and micro-arcsecond spectro-astrometry capabilities in the K-band (2.0-\SI{2.4}{\micro\meter}). 
The instrument allows one to perform phase-referenced observations by using two objects that can be observed simultaneously, one for the Fringe-Tracker (FT) channel that compensates in real-time for the atmospheric piston, and one for the science channel (SC) for long exposures. GRAVITY has delivered ground-breaking results, from the experimental test of General Relativity in the Galactic Center and the study of the environment of the Sgr A* black-hole down to the innermost circular orbit scales \citep{Gravity2018_gcredshift, Gravity2018_gcisco, Gravity2019_gcdistance, Gravity2020_gcprecession, Gravity2022_gcmass}, to the first spatially resolved observations of a broad-line region (BLR) in an active galactic nucleus \citep[AGN,][]{Gravity2018_qso}, the first detection and characterization of exoplanets using interferometry \citep{GCollab2019_hr8799, GCollab2020_betapic, Nowak2020, Lacour2021}, the first resolved image of a gravitational microlens \citep{Dong2019}, and spatially-resolved surveys of the inner disks of young stars and their magnetospheric accretion regions \citep{Gravity2019_yso,Gravity2020_ysotwhya,Gravity2021_ysottau,Gravity2023_ysottau,Gravity2024_ysoscra}.

\subsection{From GRAVITY to GRAVITY+}
\label{subsec:gravityplus}

The goal of the GRAVITY+ project is to increase the sensitivity, sky-coverage, and high-contrast capability of GRAVITY and of the VLTI infrastructure by several orders of magnitude \citep{Messenger2022}, with the primary objectives being to (1) measure the dynamical masses of Super Massive Black Holes in AGNs across Cosmic Time by resolving their BLR (2) characterize the physics of exoplanets at astronomical-unit separations by directly detecting their continuum emission or albedo and spectral signature, and (3) constrain the spin of SgrA* in the Galactic Center by unveiling and following faint objects on close-in orbit within the potential well of the black hole.

The GRAVITY+ project was first presented in 2019 \citep{Eisenhauer2019,2019Msngr.177...67M}. In 2021, its phased implementation started with the GRAVITY Wide mode \citep[hereafter G-Wide,][]{GWide2022}, which enabled for the first time wide-field observations with GRAVITY. This upgrade leverages the dual feed capability baked into VLTI since its conception and developed by the former PRIMA project \citep{10.1117/12.2054723}. Before this upgrade, the maximum FT-SC separation was limited to 2 arcseconds on the UTs and 4 arcsecond on the ATs. This requirement is overcome by G-Wide, which enlarges the maximum separation between the FT and SC to 30 arcseconds (both for ATs and UTs) by using the star separator (STS) at the Coud\'e focus of the telescopes \citep{10.1117/12.618258}. The much larger area in which to pick up a bright star suitable for fringe-tracking is virtually limited by the isopistonic angle, thus increasing the field-of-view and the sample of observable objects. This instrumental upgrade was necessary for the observations of faint sources with interferometry, in particular extragalactic sources, which are too faint to allow for fringe-tracking. The G-Wide upgrade enabled the dynamical mass up of a super massive black hole (SMBH) to be measured to a redshift $z\sim2$ for the first time \citep{GCollab2024_z2}. 

The implementation of GRAVITY+ continued with the completion of an upgrade of the GRAVITY metrology, the introduction of the so-called FAINT mode in order to remove the noise of the metrology laser \citep{Widmann2022}, an improvement of global throughput by combining the beam compressor and the differential delay lines in the VLTI laboratory \citep{Fabricius2024}, and the upgrade of the GRAVITY fringe-tracker \citep{Nowak2024_ft}. These upgrades will soon be complemented by the installation of high-precision injection optics for astrometry in GRAVITY and a high-resolution grism $R\sim15,000$ \citep{Messenger2022}, and possibly an improved vibration control for VLTI \citep{Bigioli2022,Laugier2024}.

The deployment of GRAVITY+ has put the system in a situation where most cases are limited by the performance of the MACAO system. On the one hand, observations of faint objects with the wide-field mode are in practice limited to a handful of targets because the limiting magnitude of the fringe-tracker ($\Kmag\lesssim12$) corresponds to objects that are too faint in the visible for decent AO correction ($\Vmag\lesssim13$). On the other hand, observations of ever more contrasted exoplanets require a higher-order AO correction to maximize the collected flux of the planet, while reducing the speckle and photon noises coming from the nearby bright star \citep{CRPHYS_2023__24_S2_115_0}. As explained above in this introduction, these barriers can only be lifted by the deployment of an AO facility that enables both NGS and LGS operations. This is exactly the purpose of the GPAO system, developed in the framework of the GRAVITY+ project.

\section{Instrument design}
\label{sec:design}

\begin{figure*}[ht]
\centering
\includegraphics[width=\textwidth]{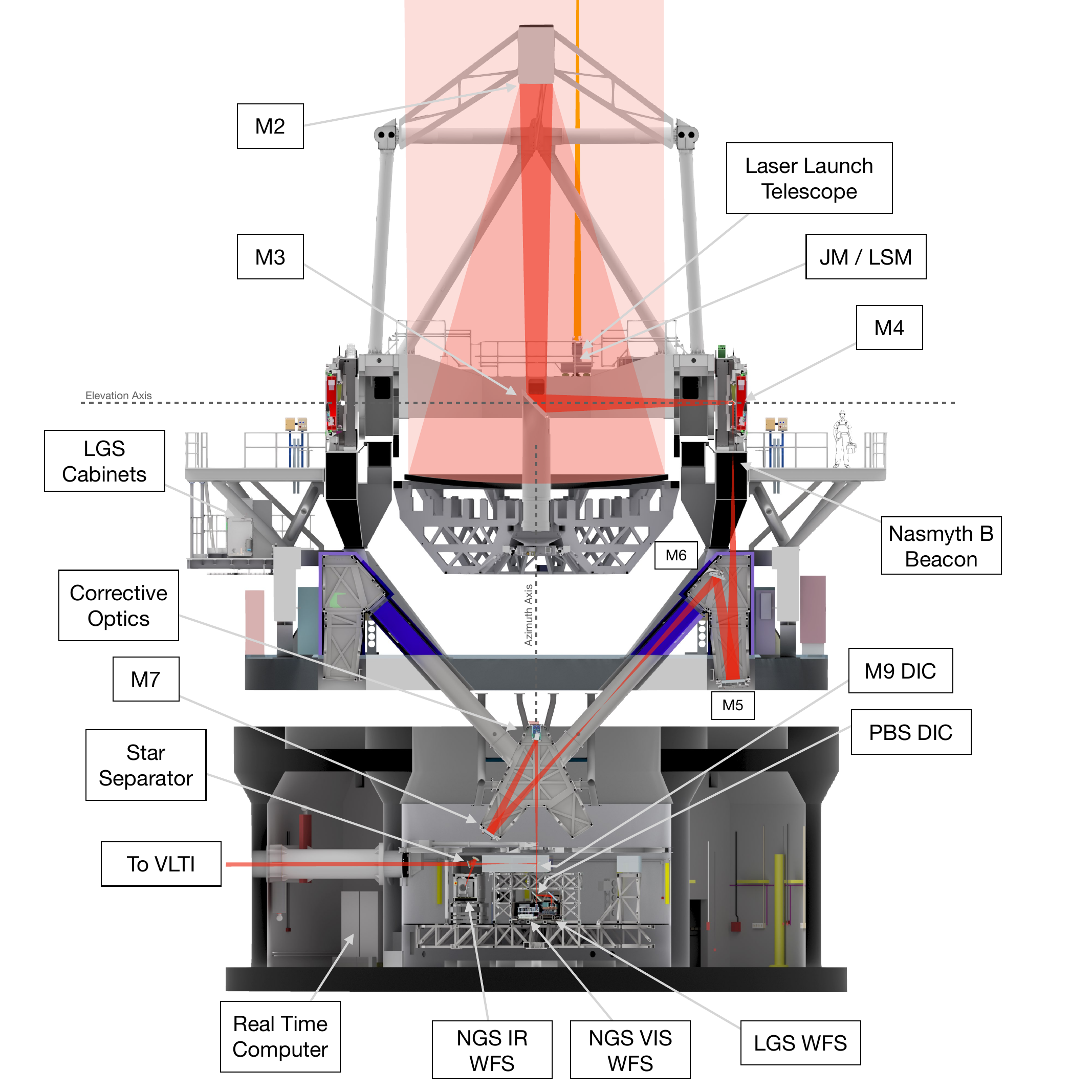}
\caption{Location of the subsystems of GPAO within the structure of the unit telescope. The beam of light is represented in light red, from the vertical incidence on M1 down to the exit toward the VLTI tunnel. Although they are not formally GPAO subsystems, the location of the M2 mirror, the elevation and azimuth axes, the Nasmyth beacon, the mirrors of the Coud\'e train, the M9 dichroic, and the star separator are also highlighted because of their particular importance for GPAO.}
\label{fig:GPAO_in_UT}
\end{figure*}

\subsection{General design guidelines and trade-offs}

The analysis resulting in the final design of GPAO inherits from: (I) the Enhanced Resolution Imager and Spectrograph \citep[ERIS,][]{2023A&A...674A.207D} for the overall error budget, the implementation of the new wavefront sensors, and the Standard Platform for Adaptive optics Real Time Applications (SPARTA) framework for the Real Time Computer (RTC) , (II) the Adaptive Optics Facility, the GRound layer Adaptive optics Assisted by Lasers and the Ground Atmospheric Layer Adaptive Corrector for Spectroscopic Imaging~\citep[AOF, GRAAL and GALACSI,][]{Stroebele2006} for the operational experience at Paranal observatory \citep{2016SPIE.9909E..2HP,2020SPIE11448E..0VH}, especially with the LGS, (III) CIAO \citep{Kendrew_2012} for the infrared wavefront sensor hardware and its operation, and (IV) the New Adaptive Optics Module for Interferometry~\citep[NAOMI,][]{2019A&A...629A..41W} for the overall interface concept with VLTI, as well as the emphasis on automatic and reliable operations. On the other hand, Appendix~\ref{appendix:ERIS} lists the differences between GPAO and these projects that have a significant impact on the design.

\subsubsection{ALPAO 43x43 deformable mirror}

The size of the deformable surface is imposed by the design of the VLTI Coud\'e train (see Fig.~\ref{fig:GPAO_in_UT}), where the pupil defined by the secondary mirror (M2) is reimaged on the mirror number 8 (M8) with a size of $98\times\SI{101}{\milli\meter}$. The voice coil technology from ALPAO\footnote{www.alpao.fr} was identified as the most cost-effective solution to achieve the required stroke (\SI{>15}{\micro\meter} peak-to-peak wavefront for $3\times3$ actuators), speed (\SI{>800}{\hertz}) and the necessary pupil size (\SI{101}{\milli\meter}).

The actuator density corresponds to an extreme-AO system (effective pitch of about \SI{20}{\centi\meter} on-sky), because (1) there is no trade-off between a high-order system and a sensitive system thanks to the LGS-assisted mode, (2) there is no trade-off between the number of actuators and the stroke or the speed in the ALPAO technology, and (3) it enables shorter wavelengths that, even if outside the immediate scope of GRAVITY+, are of interest for the long-term future of VLTI.

A prototype deformable mirror (DM) was used from October 2022 to August 2023. Two main conclusions could be drawn. First, the ALPAO standard technology (polymer springs) satisfies the requirement in term of actuator gain stability and homogeneity. Considering the risk inherent to any new technology, the mitigation plan of using innovative metallic springs for the final DMs was dropped. Secondly, the edges of the pupil were poorly controlled by the $41\times41$ \SI{2.5}{\milli\meter} pattern. The design was thus upgraded to $43\times43$ with a custom pitch of \SI{2.62}{\milli\meter}. This is the design for the final six GPAO DMs (four in operation and two spares).

\subsubsection{OCAM2 camera and frame grabber}

The Spectro-Polarimetic High contrast imager for Exopla{}nets REsearch \citep[SPHERE,][]{Beuzit2019}, GALACSI, GRAAL and ERIS use a camera internally developed by the European Southern Observatory ESO based on the EMCCD220 chip from e2v. GPAO instead uses the commercial OCAM2 camera from First Light Imaging\footnote{www.first-light-imaging.com} based on the same chip, with comparable performances (sub-electron readout noise, \SI{>1}{\kilo\hertz} full frame rate). However, the video protocol of this camera (Camera Link Full) is not supported in the ESO standard, and the physical distance between the wavefront sensor (WFS) location and the RTC cabinets renders a conversion to optical fiber necessary. GPAO uses a commercially available frame grabber in order to execute the protocol and media conversions (see Sec.~\ref{sec:real_time_calculator}). The stability of the data stream and the additional latency in the frame grabber were validated with the first camera, delivered in 2021.

\subsubsection{Shack Hartmann versus pyramid wavefront sensors}

The Pyramid sensor may appear as a more promising solution than a Shack-Hartmann (SH) for GPAO because of its improved sensitivity to the low orders, which are critical to inject inside a single-mode fiber or reach high contrast at a small inner working angle. Moreover, assuming a detector with sub-electron readout noise, a Pyramid sensor allows the number of measured modes to be reduced without modifying the optical setup (e.g., when the NGS WFS is used as a low-order sensor in LGS mode).

For the NGS VIS WFS, the Pyramid sensor has been discarded for three reasons. First, the correction of the large amplitude non-common path aberrations (NCPA) created by the downstream VLTI optical train challenges the linearity of the Pyramid sensor. Second, the selected DM technology suffers from gain instabilities and creep (with typical timescale of several seconds to days). We adopted the precautionary principle to associate this possibly nonlinear actuator with a linear, reproducible sensor so that a recalibration is always possible. And third, the development in the RTC to support the Pyramid with respect to the baseline ERIS algorithms were in contradiction with the timeline of the GPAO project. We also considered the lack of operational experience with the Pyramid sensor within ESO, while the SH had been operated with a high level of reliability for many years, which is critical for a complex system like VLTI.

For the LGS WFS, the Pyramid sensor has been discarded because the critical low orders are sensed by the NGS sensor anyway. Furthermore the behavior of a Pyramid WFS with the LGS elongated spot is not yet completely understood. 

\subsubsection{Number of measured modes}

The philosophy motivating the design is the following: the GPAO NGS mode is optimized for robust peak performances on bright stars, the GPAO LGS mode is optimized for robust performances toward the faint end. On the one hand, considering the DM characteristics, the maximum useful repetition rate of the AO loop is \SI{2}{\kilo\hertz}, which is matched by the OCAM2 camera in full frame. Therefore, there is no trade-off between the number of pixels to read and the speed. On the other hand, the RTC prototyping activity predicted a maximum loop rate of \SI{2}{\kilo\hertz} for an ERIS-like system ($\approx40\times40$ sensing, $\approx1200$\,actuators). So, there was no trade-off between the number of sub-apertures and the speed. Consequently, the baseline design for the GPAO NGS mode is to control up to 900 modes with a $40\times40$ SH.

The LGS WFS controls up to 500 modes with a $30\times30$ SH. With respect to a $40\times40$, the $30\times30$ provides improved robustness against situation of low-sodium density. A $20\times20$, while obviously even more robust to lower flux return, would have significantly degraded performances in the case of average seeing and hampered the perspective of shorter wavelengths at the VLTI.

\subsubsection{Piston-free operation}

An AO system for an interferometer will provide a piston-free correction. There are subtleties in this definition (e.g., mean defined over pupil or pupil apodized by the single mode fiber of the instrument) but the analysis is identical at first order. The piston properties of the ALPAO mirror technology were investigated by the NAOMI project \citep{2019A&A...629A..41W}. First, this study demonstrated that it is possible to construct a piston-free modal basis, without the need for a position sensor \citep{2018SPIE10703E..71L}. Secondly, this study showed that the stroke of piston-free modes is only slightly reduced  compared to their optimal counter-part including piston, and only for low-order radially symmetric modes (focus, spherical, and higher order spherical). Altogether, we concluded that the classical requirements on DM manufacturing (stability, calibrability) and on the RTC design (modal control) fit the need of GPAO.

\subsubsection{Integration of CIAO}
\label{sec:integration_of_ciao}

CIAO is the $9\times9$ SH WFS providing off-axis infrared (IR) wavefront sensing for the observations of the Galactic Center with GRAVITY \citep{Scheithauer2016}. It is fed by the second beam getting out of the star separator (STS) at the Coud\'e focus of the UTs. The CIAO RTC is built on the SPARTA-light platform. It controls 40 modes, to match the degree of freedom of the MACAO deformable mirror.

In order to maintain the IR wavefront sensing at VLTI, the CIAO hardware is fully integrated as an additional sensor of the GPAO system. The opto-mechanical hardware and the local controllers remained unmodified. The CIAO RTC, the workstations and the high level control software have been decommissioned. This choice was motivated for its opportunity of combining the CIAO infrared sensor with the new LGS sensor for the Galactic Center. The nearby GCIRS7 star (\Kmag=6.5 at 6\arcsec{}) is used for sensing the tip-tilt and focus (with the possibility for up to 44 modes) in the IR while the LGS WFS senses the higher order modes. Without IR sensing, the closest visible star is fainter and further away (\Vmag=14 at 15\arcsec{}). This approach also strongly simplifies the operations by merging all sensors (VIS, IR) into a common AO system for VLTI/UT.

\subsubsection{Zonal versus modal control}

Fundamentally, the control in GPAO is modal, in the sense that the control space is determined by a subset of predefined Karhunen-Lo\`eve modes (KL, see Sect.~\ref{sec:KL_modes}), and not by a subset of the eigenvectors of the WFS+DM system. This classical approach in modern AO delivers improved performances by focusing on the modes with highest turbulence power. The control can be "explicitly modal" (a different temporal filter is applied on each mode) or "implicitly modal" (the modal space is defined by the row-space of the control matrix, complement of its kernel, and the same temporal filter is applied to all elements of the command vector). Systems exist in both approaches.

The explicitly modal approach allows the temporal filters to be tuned differently on each mode, for instance to reject known vibrations on a given mode, or to implement large leaks on high-order modes. But its implementation in SPARTA suffers from an increased latency because of a second, post-filter matrix multiplication (modes to DM commands). Considering the risk on the latency and the necessary development time, it was decided to implement the high-order controllers of GPAO as implicitly modal. Ultimately, modal leaks could still be implemented in these implicitly modal controllers, as described in Appendix~\ref{appendix:saturation}. 

\begin{table*}[t!]
\caption{Parameters of GPAO wavefront sensors associated with their corresponding AO mode.}
\label{tab:WFS_parameters} 
\centering
\begin{tabular}{l c c c c c c}
\hline\hline
  \text{Parameter}  &  LGS & NGS VIS HO & NGS VIS LO & NGS IR HO & NGS IR LO \\\hline
   \\[-1.7ex] \vspace{1mm}
   Camera & OCAM2 & \multicolumn{2}{c}{OCAM2} & \multicolumn{2}{c}{Saphira} \\
  Nb. of pixel read & $240\times240$ & \multicolumn{2}{c}{$240\times240$} & \multicolumn{2}{c}{$72\times72$} \\
  \cmidrule(lr){3-4}
  Pattern of subap & $30\times30$ & $40\times40$ &  $4\times4$ & \multicolumn{2}{c}{$9\times9$} \vspace{1mm}\\
  Nb. of subap & 704 & 1240 &  12 & \multicolumn{2}{c}{68} \vspace{1mm}\\
  Plate scale [\arcsec/pix]& 0.71 & 0.42 & 0.21 & \multicolumn{2}{c}{0.51}\\
  \cmidrule(lr){5-6}
  Frame rate [Hz] & 1000 & 1000,500,250,100 & 500,250,100 & 1000,500,200,100 & 500,200,100 \vspace{1mm}\\
  Controlled KL modes & 4-500 & 500,200 & 3 & 50 & 3 \vspace{1mm}\\
  Modal controller & implicit & implicit & explicit & implicit & explicit \vspace{1mm}\\\hline
  \\[-2ex] 
   \text{is active in mode:}\\
   NGS\_VIS &   & \checkmark &   &   &  \\
   NGS\_IR  &   &   &   & \checkmark &  \\
   LGS\_VIS & \checkmark &   & \checkmark &   &  \\
   LGS\_IR  & \checkmark &   &   &   & \checkmark\\
   \hline   
\end{tabular}
\tablefoot{In terms of hardware, the NGS VIS HO/LO wavefront sensors only differ by the setting of the NHILO lenslet. In terms of hardware, the NGS IR HO/LO wavefront sensors are identical.}
\end{table*}

\subsubsection{Software derotations}
\label{sec:software_derot}

GPAO is unusual in the number of rotations inside the system. The photometric pupil and its spiders (M2), as well as the Jitter mirror of the Laser Launch Telescope, rotate with respect to the WFS by the angle \emph{Azimuth - Elevation}, which is constantly changing while the telescope is tracking. The DM rotates with respect to the WFS by the \emph{Azimuth} angle, which is also constantly changing. We decided to avoid any mechanical derotation inside the WFS, meaning that both the photometric pupil and the DM pattern rotate on the WFS. This strategy rigidifies the entire ensemble of M9 + Plano Beam Splitter (PBS) + WFSs + STS + VLTI into a common, nonrotating frame (see Fig.~\ref{fig:GPAO_in_UT} and discussion in Sect.~\ref{sec:software_derot} and Fig.~\ref{fig:GPAO_control}). Doing so significantly simplifies the geometrical modeling of the instrument and the management of the NCPA.

Consequently, GPAO has to implement a numerical derotation. The strategy implemented in the former CIAO (measuring the DM/WFS interaction matrix at many Azimuth angles, defining a subset of system modes properly seen at all angles, and precomputing a bank of control matrices) would be extremely time-consuming in calibration considering the number of modes in GPAO and the increased sensitivity to the system angle. The strategy implemented in NAOMI (rotating the controlled Zernikes) is not possible in GPAO because the control space (KL modes) does not define a rotation operator. Instead, GPAO relies on pseudo-synthetic interaction matrix (PSIM, see Sect.~\ref{sec:PSIM}) created with a geometric model of the instrument. The angle between the DM and the WFS is one of the parameters of this geometric model. The control matrix is recomputed on-the-fly based on the actual Azimuth angle. The highest update rates (\SI{5}{\second}) allow the meridian to be crossed at a zenithal distance of 5\,deg while controlling 900 KL modes.

\subsubsection{Patrol fields}

For the NGS WFS, it makes sense for the patrol field to be larger than the isokinetic angle in the longest wavelength used by VLTI (\SI{10}{\micro\meter}), in good seeing conditions. In practice, this means covering the entire \SI{1}{arcmin} (radius) telecentric field of view at the Coud\'e focus of the UTs.

For the LGS WFS, we also decided to patrol the entire Coud\'e field of view. This made it possible to put the LGS star on any of the two objects transmitted to VLTI (one being the fringe-tracking star, the other being the science object). It also allows small pointing offsets to be added to the LGS spot, fixed in pupil orientation, as a mitigation to potential Rayleigh contamination.

\subsection{Description of subsystems}
\subsubsection{Implementation in the Unit Telescopes}

Figure \ref{fig:GPAO_in_UT} highlights the location of the main subsystems of GPAO within the structure of the Unit Telescope. Relevant optical elements are the secondary mirror M2 that defines the aperture stop of the telescope, the DM located in M8 and conjugated with M2, the dichroic beam splitter M9 that reflects the infrared part $\gt \SI{1}{\micro\meter}$ for the VLTI/STS/CIAO and transmits the optical light to the visible and laser WFSs, and the telecentric doublet lens located just below M9 in the transmitted Coud\'e focus. This lens acts as a (quasi) field lens 
and provides a telecentric field plane of \SI{1.5}{arcmin} at f/47, the so called patrol field. The M9 dichroic being in a convergent beam close to focus, it generates field-dependent aberrations and pupil distortions in the transmitted beam.

\subsubsection{Wavefront sensors}
\label{sec:wfs}

The new wavefront sensors of GPAO enable NGS VIS and LGS operations \citep{2024SPIEBou...WFS}. Their design is based on the AO module of ERIS \citep{2018SPIE10703E..03R}, adapted from a f/13.6 beam to a f/47 beam. The WFS are located at the Coud\'e, below the VLTI STS and M9 dichroic (Fig. \ref{fig:GPAO_in_UT}). Inside the WFS unit, the Periscope Beam Splitter (PBS, a plano-parallel dichroic) reflects the light of the sodium laser line around \SI{589.195}{\nano\meter} toward the LGS WFS, and transmits the broad band natural light ($600-\SI{1000}{\nano\meter}$) toward the NGS VIS WFS. Each of the two wavefront sensors are installed on a translation XY-stage that cover the full patrol field of \SI{2}{arcmin}. The parameters of the WFS corresponding to the respective AO modes are indicated in Table \ref{tab:WFS_parameters}.

\paragraph{NGS VIS WFS} The natural guide star visible WFS is a Shack-Hartmann that can be configured either as a high-order (HO, $40\times40$) sensor or as a low-order (LO, $4\times4$) sensor. It is composed of eight main subsystems. (i)~The \component{atmospheric dispersion compensator} (ADC) is composed of a pair of Amici prisms, and designed for a maximum zenith angle of $70^{\circ}$. (ii)~The \component{technical charge couple device} (TCCD), uses the blue part of the spectrum $<400-\SI{570}{\nano\meter}$ ($\mathrm{R} > 95\%$) reflected from an internal dichroic (DCR). It allows either the pupil or the field to be imaged using a moving lens, and can be used as an acquisition camera with 20\arcsec{} field of view. (iii)~The \component{NGS pupil steering mirror} (NPSM) adjusts the position of the pupil with a piezo-driven mirror, located close to an intermediate image plane with f/20. (iv)~The \component{NGS field diaphragm} (NDIA), located in the image plane, limits the field-of-view of the subapertures to avoid cross-contamination. (v)~The \component{filter wheel} is a set of neutral densities (1/10 and 1/100), a He-Ne notch filter, a Sodium notch filter, and a long-pass $>\SI{800}{\nano\meter}$ filter for best performances on bright stars. (vi)~The \component{NGS derotator} (NROT), made of FK5 prism. In normal operation, this device is fixed. (vii)~The \component{HO/LO} translation stage switches between either the HO or LO lenslet arrays. (viii)~The \component{OCAM2 camera} measures the output spots, with a full detector size of $240\times240$ pixels. The electron-multiplication (EM) gain enabled by the EMCCD ranges from 1 to 1000. The overall bandpass of this wavefront sensor, including the Coud\'e train, matches the \Grp{} bandpass \citep{2021A&A...649A...1G}.

\paragraph{LGS WFS} The laser guide star WFS is a Shack-Hartmann with a fixed configuration. The beam is converted from f/47 to f/12 in order to adjust for the large focusing range. It is composed of 4 main elements. (i) The \component{laser focus control stage} (LFOC), which allows the focus to be adjusted from \SI{80}{\kilo\meter} to infinity. (ii) The \component{LGS derotator} (LROT) is kept fixed in normal operation. (iii) The \component{LGS pupil steering mirror} (LPSM), located close to the image plane, steers the pupil based on the same principle as in the NGS. (iv) The \component{$30\times30$ lenslet array} is fixed inside the \component{OCAM2 camera}. When observing in LGS mode, the NGS and LGS camera are synchronized by a physical link that distributes a common clock.

\paragraph{Calibration Unit}
The calibration unit of the WFS subsystem consists of two multi-mode fibers located in the respective focal point of the NGS VIS and of the LGS wavefront sensors. The fibers are fed with a broadband halogen lamp that covers the optical/near-IR spectrum ($400-\SI{1000}{\nano\meter}$). The calibration unit is moved in and out of the optical path by a linear translation stage. The NGS IR wavefront sensor does not a have calibration unit.

\subsubsection{Corrective optics and deformable mirror}
\label{sec:co_and_dm}

The corrective optics (CO) assembly is located in the M8 position of the UT. It is composed of a motorized gimbal tip-tilt mount that hosts the deformable mirror itself.

The ALPAO Deformable Mirror (DM) features 1432 actuators on a square \SI{2.62}{\milli\meter} pitch pattern. The stroke of pushing a group of $3\times3$ actuators is typically \SI{20}{\micro\meter} peak-to-peak wavefront, with respect to the zero position (that is: this stroke is achievable in both directions). The first resonance is \SI{1000}{\hertz}. The DM is controlled by two standard Drive Electronics of ALPAO, each featuring two Ethernet connections. One is dedicated to house keeping and monitoring, and the other to the high speed data link (\SI{1}{Gbit/\second}). The DM is equipped with cooling plates on two of its lateral faces. The flow of coolant is regulated to maintain the DM internal temperature to less than 3\,deg above the ambient temperature at the location of the DM.

The center of the DM (actuator 716, starting 0) is located on the Azimuth axis of the telescope with an accuracy better than \SI{0.5}{\milli\meter}. The DM is mounted in a quasi-static gimbal mount (QSM) to provide a large range (2\,deg-mechanical, $>2\,$arcmin-sky), slow tip-tilt with the vertex in the center of the DM surface. The actuation is absolutely encoded with a resolution of 10\,mas-sky. This actuation serves to steer the Nasmyth Beacon anywhere in the Coud\'e field of the UT, and to offload the DM tip-tilt when doing calibration. When observing on sky, the QSM is kept static at the position corresponding to the optical axis of the telescope.

The electronic cabinet of the Corrective Optics is located inside the inner track of the UT to keep the distance to the DM shorter than \SI{3}{\meter}. This cabinet embarks the two DM electronic racks (each driving $\sim700$ actuators), the local controllers of the motors of the gimbal motion, and a low-latency Ethernet switch used to convert the incoming high speed and housekeeping networks (transported over fibers) into copper cables.

\subsubsection{Laser guide stars}
\label{sec:laser_guide_star}

GPAO will be operated with the LGS to be installed on UT1-2-3 \citep{2024SPIE13096E..9AB} and with LGS\#1 installed on UT4 \citep{aot-2014-0025}. The laser sources of UT1-2-3 are essentially identical to those used in the UT4 Adaptive Optics Facility \citep{2011aoel.confE..52K} and those that will be used on the ELT. The LGS laser source feeds a laser projection subunit (LPS) mounted on the side of the UT primary mirror. The LPS of UT1-2-3 are identical to those used on the ELT \citep{10.1117/12.3020152} and are comparable in performance to the one of UT4. The LPS is basically a reversed telescope equipped with a fast Jitter Mirror (JM) and a slow Laser Steering Mirror (LSM) used to control the beam direction. Each UT is equipped with its own camera-based aircraft avoidance system to prevent beam collisions with planes.

\subsubsection{Real time computer}
\label{sec:real_time_calculator}

The real time computer (RTC) is made of two components. First, the hard RTC (HRTC) is responsible of the main low-latency control loop. Second, the soft RTC (SRTC) is in charge of monitoring and driving the HRTC with high level, but not real-time, tasks. GPAO reuses and improves many features of the ERIS RTC \citep{2018SPIE10707E..1HB}, the AOF RTC \citep{2012SPIE.8447E..2DK} (e.g., the real time calculations blocks) and of the NAOMI RTC (e.g., the chopping and the saturation management, and the communication with the ALPAO deformable mirrors).

\paragraph{HRTC} The HRTC is based on the SPARTA-upgrade platform from ESO \citep{Shchekaturov:23_SPARTA_upgrade}, which replaces the SPARTA RTC box by a single workstation. The HRTC receives the pixels of the OCAM2 cameras via Pleora iPort CL-Ten frame grabbers in GigE-Vision (10 Gigabit Ethernet, GbE). The HRTC receives the pixels of the NGS IR Saphira detector using the native serial front panel data port (sFPDP). It sends command to the DM using the ALPAO HSDL protocol through two 1\,GbE links (one per electronic). The HRTC also sends commands to the JM and LSM mirrors in ESO MUDPI format through a 10\,GbE link. The hardware chosen for the HRTC is the Dell PowerEdge R7525 server, equipped with additional Intel X710-T4 Network Interface Cards and one New Wave DV v5051 sFPDP Card. The system is configured to run as deterministic as possible (diskless, load only required interfaces, confine the kernel). It takes advantage of its multi-CPU architecture, of the principle of Interrupt Request affinity and of the NUMA architecture. The HRTC use non-blocking communications and shared memory areas to exchange intermediate results so that the data can be processed in parallel to their reception as much as possible.

\paragraph{SRTC} The SRTC is the interface between the HRTC and the outside world. It catches the HRTC real-time stream, analyzes performance parameters, triggers alarms, and updates the matrices and references vectors. The SRTC mostly interacts with the Observation Software (OS) and its secondary loops (see Appendix~\ref{app:SL}), to which it sends measurements (e.g., pupil position, mean tip-tilt), and from which it receives updates and commands (e.g., new rotation angles, trigger to recompute the control matrix, new NCPA vector). In terms of hardware, the SRTC is implemented as a standard SPARTA cluster composed of several workstations connected via a 10\,GbE switch. 

The SRTC implements different modules that were adapted or improved to meet GPAO needs. (i)~The \component{PSIM module} implements the synthetic model of the system used to generate the control matrices (see Sect.~\ref{sec:PSIM}). (ii)~The \component{misregistration monitor} fits the PSIM parameters, either from calibration or inferred from the AO telemetry, as described by~\cite{2012SPIE.8447E..2CB}. (iii)~The \component{shift monitor} is dedicated to lateral error estimation. It implements innovative methods \citep{2024A&A...687A.157B, 2024SPIE13097E..0TB} with a fast and robust, but perturbative, solution for the automated acquisition, see Sect.~\ref{sec:automated_acquisition}, and a noninvasive approach for closed loop operations, see Sect.~\ref{sec:misregistration}. (iv)~The \component{spot monitor} uses a global inverse problem approach to robustly fits the spot parameters jointly through all the SH-WFS subapertures: spot flux, spot position, spot size in NGS, sodium layer thickness in LGS, elongation, and orientation of natural extended objects. The robustness of the approach makes it an essential tool for the automated acquisition (see star detection, star centering, magnitude estimation in Sect.~\ref{sec:automated_acquisition}). This module also computes the weight maps for the weighted center of gravity used on faint or extended objects. (v)~The \component{loop monitor} analyzes the AO loop telemetry to compute average offloads (focus, tip-tilt) or to raise alarms (DM clips or saturation, under-illumination, over-illumination) to inform the OS which is in charge to open or not the loops. (vi)~The \component{atmosphere performance monitor} estimates the atmospheric conditions ($r_0$, $L_0$, seeing, Strehl, ...) based on the pseudo-open loop slopes computed from the AO telemetry. It is directly inherited from AOF. A new wind speed estimator has been integrated as well, based on a frozen flow multilayer model. (vii)~Finally, the \component{pupil monitor} follows the lateral position of the photometric pupil, with the possibility to finely fit the pupil magnification and relative angle.

\begin{figure*}[t!]
\centering
\includegraphics[width=\textwidth]{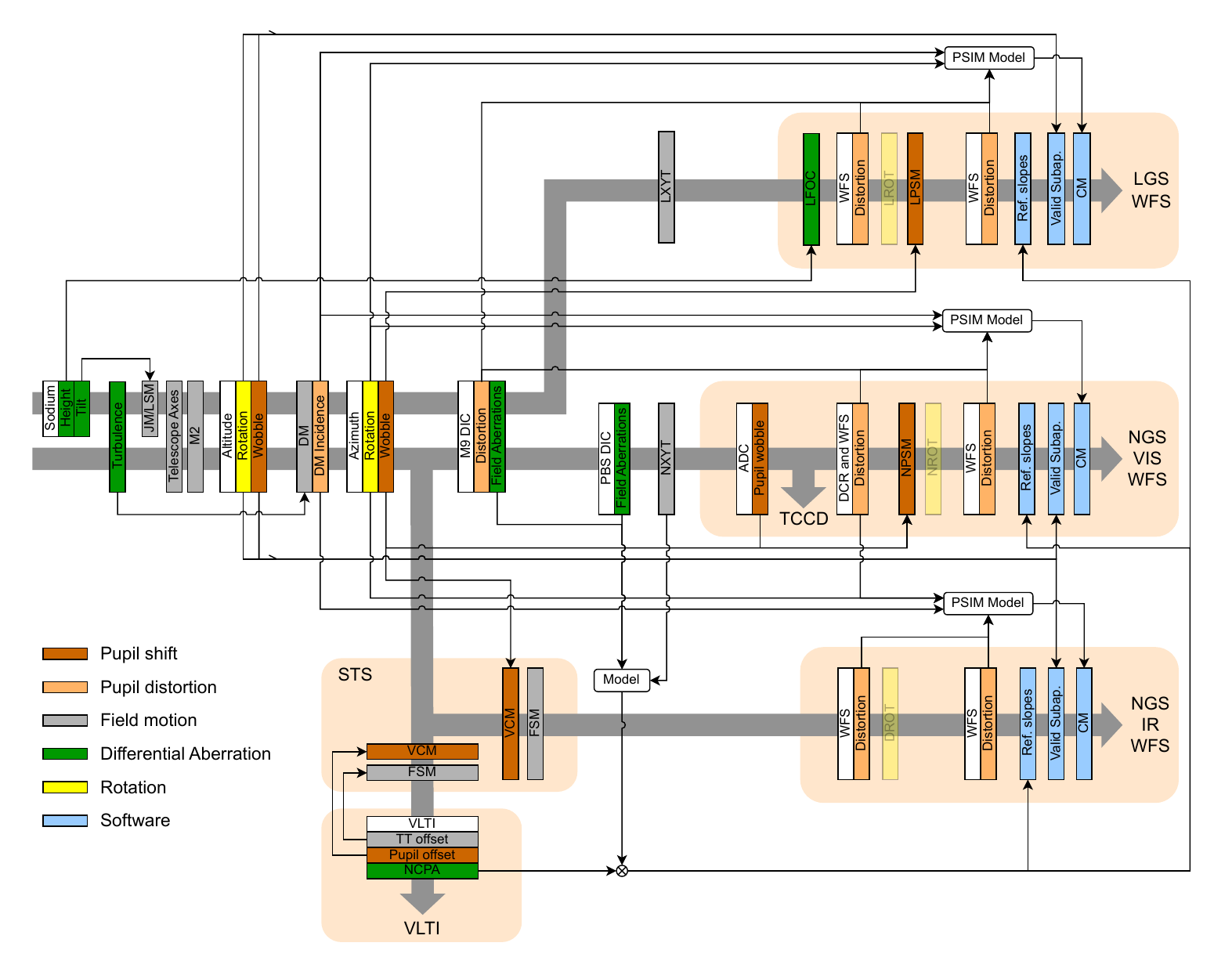}
\caption{Control scheme of GPAO. Disturbances and corrections are represented but sensing is not represented. The derotators LROT, NROT, and DROT are shaded to express that the hardware is physically present but that they are kept at a fixed angle.}
\label{fig:GPAO_control}
\end{figure*}

\subsubsection{Control software}
\label{sec:control_software}

The control software of GPAO is independent for each telescope, both in hardware and software. Each GPAO has two workstations running the framework of the VLT Software. The GPAO Instrument Workstation hosts the observation software (OS) and the instrument control software (ICS). A Detector Control Workstation (DCS) acts as command handler layer toward the camera.

The ICS controls and supervises all GPAO devices, including all devices of the former CIAO instrument. It communicates to the hardware via local controllers. It is the only interface for piloting the motion of axes. It reads the sensor telemetry and house keeping, including those from the DM.

The OS coordinates the activities of the ICS and the RTC. It collects the states and logs of all subsystems and constructs the main state and logs of GPAO. It executes the acquisition sequence. It runs the secondary loops described in Appendix~\ref{app:SL}. It prepares telemetry data to be archived. Toward the outside, it receives commands from the VLTI Supervisor, it sends commands to the Telescope Control Software, to the star separator of the VLTI and to the Multi Laser Guide Star Facility of the Unit Telescope.

\subsection{Integration and testing of subsystems}

Considering its importance for GRAVITY+, and the maturity of extreme-AO and LGS-assisted AO, the Gravity Plus Adaptive Optics has been implemented with a fast schedule. The design was finalized in 2022. The complete systems GPAO1 and GPAO2 were integrated and tested from the end of 2022 to early 2024, in a dedicated test bench that reproduced the telescope interfaces \citep{2024SPIE13095E..20M}. The remaining systems GPAO3 and GPAO4 were assembled and verified at subsystems level in the different institutes. The four GPAOs were installed at the Paranal observatory in mid-2024 and regular operations of the NGS-mode started in December 2024.

\subsection{Control strategy}
\label{sec:control_strategy}

\subsubsection{Karhunen-Lo\`eve modal basis}
\label{sec:KL_modes}

The modal basis of GPAO uses approximate Karhunen-Lo\`eve (KL) modes defined in DM command space as described in Appendix~B of \cite{2022A&A...658A..49B}. In practice, the five first modes are enforced to be exact Zernikes (tip, tilt, focus and the two astigmatisms) and the piston is removed from the basis. The controlled pupil is restricted to an annulus extending from 13.95\% to 100\% of the pupil diameter. This defines 1125 active actuators, thus 1125 independent modes, and 307 passive actuators, the latter being controlled by a linear combination of the former. The exact same modal basis is used for all wavefront sensors in all configurations (NGS\_VIS, NGS\_IR, LGS\_VIS, LGS\_IR).

The number of inverted modes (e.g. controlled modes) changes depending on the wavefront sensor in use, when building the control matrix. The values can be found in Table~\ref{tab:WFS_parameters}. For the high-order natural guide star wavefront sensors, the number of controlled modes is reduced when approaching the limiting magnitude. These numbers have been selected to maximize the robustness of the system while still matching the requirement in terms of Strehl ratio in the K-band under median seeing conditions. Controlling more modes will be explored in the next years, especially in the context of future J-band instruments \citep[e.g.][]{2022SPIE12183E..1SK, Lacour2025}. When using the $40\times40$ WFS, the system has significant margin to control possibly up to 950 modes. 

\subsubsection{NGS/LGS combined control}

In NGS mode, all controlled KL modes are sensed by the NGS WFS and corrected by the HRTC with the DM. In LGS mode, the tip-tilt and focus are sensed with the NGS LO WFS and the higher order KL modes are sensed with the LGS WFS. All are corrected with the DM. The filtering of controlled KL modes in each pipeline is entirely defined by the construction of its control matrix.  A detailed view of the real-time dataflow is  presented in Appendix~\ref{appendix:saturation}.

Because of the up/down propagation through turbulence, and because of the slowly varying distance of the Sodium layer, the tip-tilt and focus of the laser spot in the LGS WFS also needs to be controlled. The fast tip-tilt is corrected by the HRTC with the Jitter Mirror (JM) of the laser launch telescope. The JM offloads to the Laser Steering Mirror (LSM). The slow drift in focus of the laser spot is corrected by a Secondary Loop (SL) on the OS: it reads the averaged focus on the LGS WFS from the SRTC, and acts on the LFOC device accordingly.

\subsubsection{Tip-tilt offload}
\label{sec:tiptilt_offload}

Offloading the tip-tilt modes from the DM reduces the overall stroke budget and cancels the fitting error of the tip-tilt modes. A dynamical Gimbal mount for a DM weighing \SI{5}{\kilo\gram} was considered too complex for GPAO. Instead, the initial GPAO design considered offloading the tip-tilt into the M2 itself, using the available "rapid-guiding" link of the VLT (running at \SI{100}{\hertz}) with a small gain ensuring an offload bandwidth of a few Hertz.

When configured in rapid-guiding, the Dornier M2 of UT1-2-3 delivers tip-tilt residuals of \SI{20}{\milli as}-rms. This is satisfactory for seeing-limited observations, but needs to be further reduced for AO-assisted systems. Unfortunately, preliminary tests during commissioning confirmed that most of the power of this noise reaches up to \SI{100}{\hertz}, exactly in the amplification bandwidth of GPAO (see Fig.~\ref{fig:GPAO_TF}). In the meantime, the stroke of the DMs alone was found sufficient to observe up to a seeing above 1.5\arcsec{}. The rapid offload to M2 was therefore discarded and replaced by a slower offload to the telescope axes, every \SI{3}{\second}, with a gain of $\approx0.5$. All the communication and control mechanisms remain in place, so the rapid offload could be revived if the M2 noise is reduced in the future (an upgrade of UT3 is already planned within a few years time, UT1-2 to be defined).

\subsubsection{PSIM model}
\label{sec:PSIM}

The AO control scheme in GPAO is based on pseudo-synthetic interaction matrices \citep[PSIM,][]{2006SPIE.6272E..20O}. "Synthetic" indicates that the IMs are computed from a pure numerical model of the interaction between the DM and the WFS. "Pseudo" emphasizes that this model only depends on a small number of physical parameters that are fitted on the real system.

(I)~The numerical model is based on a pure geometrical propagation of the incident wavefront: the WFS slopes are canonically given by the integrated phase differences on the lenslet sides.  The DM is modeled by a set of two-dimensional (2D) influence functions defined on a high resolution Cartesian grid. Their 2D model is non-axisymetric, and account for the inhomogeneities of the actuator gain and influence functions. This proved necessary to avoid introducing artificial fitting errors, and to properly match the desired modes.

(II)~The PSIM model includes a limited number of parameters.  Their numerical values can be found in Table~\ref{tab:PSIM_parameters}. The influence function of the DM are scaled by the DM stroke amplitude. The DM pattern is (1) stretched according to the incidence of the beam on the DM, (2) magnified by a scaling parameter to account for the magnification between the DM and the WFS, (3) rotated by an angle `azimuth - 12.984$^{\circ}$ + a reference angle' that is calibrated on the VLT beacon, and (4) stretched by two anamorphosis (at 0$^{\circ}$ and 45$^{\circ}$) to capture the pupil distortion. The distortions of the pupil at the Coud\'e focus (M9 DIC) and inside the WFS (PBS, WFS optics) can be factorized into a single, static distortion because the DROT/NROT/LROT are kept at a static position (see Fig.~\ref{fig:GPAO_control}). Note that the model does not include the lateral misregistrations because these are autocorrected in hardware in operation (see Sect.~\ref{sec:misregistration}).

\subsubsection{Lateral registration between DM and WFS}
\label{sec:misregistration}

The PSIM model introduced in Sect.~\ref{sec:PSIM} assumes that there is no lateral error between the DM and the WFS. To keep the system closed to its optimal functioning point, any alignment drift is monitored and corrected in hardware by an OS secondary loop, see Appendix ~\ref{app:SL}.

For the low-order wavefront sensors, this lateral alignment is obtained by centering the photometric pupil in the wavefront sensor. However, for the high-order sensors, the difference between the photometric pupil and the DM pupil is significant and thus the former cannot be used to track the lateral misregistration of the latter (see Appendix~\ref{appendix:ERIS}). To do so, GPAO implements an innovative approach based on the `2D+$t$' representation of DM command telemetry~\citep{2024A&A...687A.157B, 2024SPIE13097E..0TB}. In the presence of a lateral misalignment, the symmetric (cosine) part of a given spatial frequency of the command is seen with an antisymmetric (sine) small component by the WFS that leaks on the antisymmetric part of the DM command, and vice-versa. For small shifts, this coupling pattern only depends on the loop parameters (e.g., gain, latency, frequency) and is proportional to the lateral error. Contrary to other methods, this solution is noninvasive nor model-dependent~\citep{2021MNRAS.504.4274H} while the sparsity of the signal in the `2D+$t$' Fourier domain makes it fast and with a limited bias by the wind~\citep{2012SPIE.8447E..2CB, 2019VMSAI...1...12H}.

This method has been validated on-sky during the GPAO commissioning, both with the $9\times9$ NGS IR WFS and with the $40\times40$ NGS HO visible WFS. Nonetheless, in strong wind conditions, it still proved to be biased by the wind in the HO VIS mode at low frequencies. As a consequence, the auto-alignment secondary loop is disabled, without noticeable loss of performances, for loop frequencies smaller than \SI{250}{\hertz} where only 200 modes are controlled and for which a fine alignment is not necessary.

\subsubsection{Photometric pupil stability}
\label{sec:photometric_pupil_stab}

Because the DM is actively registered onto the WFS (see Sect.~\ref{sec:misregistration}), the photometric pupil (M2) is not necessarily centered. The mitigation implemented in GPAO is to zero the slopes measured on poorly illuminated subapertures. Practically, the median flux over the subapertures is measured during the acquisition phase, and a threshold of 15\% of this value is defined in the RTC. We verified on-sky that the loop remains stable even when vignetting a large fraction of the pupil by rotating the UT dome.

This mechanism proved very effective to handle the wobbling of photometric pupil with the elevation axis, which amounted to nearly three sub-apertures (in $40\times40$ mode) when installing GPAO. Still, the drawback is that a part of the wavefront, opposite to the under-illuminated subapertures, remains uncontrolled because it falls out of the DM. Considering that an offset of 10\% of the pupil diameter was unacceptable, ESO and the consortium quickly engaged into a physical realignment of the four UT telescopes. The M3 mirror was tilted to reduce the elevation wobble (see Fig.~\ref{fig:GPAO_in_UT}). The M4 mirror, closed to a focus, was adjusted to center the now stable pupil into the DM. Altogether, the photometric pupil is now always within one\,sub-aperture (in $40\times40$ mode).

\subsubsection{Chopping}

GPAO implements chopping based on asynchronous communications and absolute time stamping. The Interferometric Supervisor Software (ISS) receives the chopping parameters (amplitude and direction on sky, period, duty cycle and start time of a few seconds in the future) from the scientific instrument. The ISS forward these parameters to the M2 controller and to the GPAO OS. The GPAO OS forwards these parameters to the HRTC. Once the start time is reached, the HRTC determines the current state (``on-sky'' or ``on-science'') of the current frame purely based on the period and the absolute time. There is no other synchronization between the M2 and the HRTC rather than sharing the same time server. The HRTC freezes the loop during the ``on-sky'' phases, and resumes the loop during the ``on-science'' phases. Freezing in the RTC means that the last command is kept unmodified, without feedback from the measurements, nor accumulation of leaks or anti-windup. The SRTC monitors and recorders ignore the ``on-sky'' phases when relevant (e.g., to compute the offloads).

\subsubsection{Non-common path aberration}
\label{sec:ncpa}

As discussed above, non-common path aberrations (NCPA) inside the WFS and inside the VLTI can be factorized into a single, static aberration because the DROT/NROT/LROT are kept at a static position (see Fig.~\ref{fig:GPAO_control}). These NCPA are measured all the way to the IRIS guiding camera of VLTI \citep{2004SPIE.5491..944G, Pourre2024_phd}, using the telescope Nasmyth beacon in day time. Sequentially on each controlled mode starting with defocus, a modulation is injected into the closed AO loop at the level of the WFS slope measurements. The variance of the IRIS image, a proxy of the Strehl ratio, is computed for each recorded frame. The modulation on each mode contains a high \SI{25}{\hertz} frequency and two periods of a low \SI{0.25}{\hertz} frequency. The amplitude of the \SI{25}{\hertz} detected on IRIS reaches a minimum when the slow period crosses the NCPA offset (the Strehl maximum). Two periods are recorded, giving at least three consecutive minima and their temporal spacing is a robust estimate of the NCPA offset.

Theoretically, the previous calibration is no longer valid when the beam crosses the M9 and PBS dichroic at a different location, that is when the NGS WFS is off-axis. This field-dependent NCPA is modeled by a low-order polynomial dependence of the astigmatisms modes. The model can be calibrated on measurements made on the beacon at various off-axis positions, thanks to the steering capability of the QSM. As of today, this calibration is set to zero since no significant field-dependent NCPA has been observed in VLTI.

\section{Operations and performances}
\label{sec:op_and_perf}

\subsection{Operations}

\subsubsection{Operational modes}
\label{sec:operational_modes}

GPAO supports the four modes described in Table~\ref{tab:WFS_parameters} (NGS\_VIS, NGS\_IR, LGS\_VIS, LGS\_IR), consisting of various combination of the wavefront sensors. Seeing-limited mode (no AO correction) and seeing-enhancer mode (only LGS correction) are virtually useless for the spatially filtered instruments of VLTI, and are thus not supported.

\subsubsection{Automated acquisition}
\label{sec:automated_acquisition}

A VLTI preset (change of target) is triggered by the scientific instrument, namely PIONIER \citep{2011A&A...535A..67L}, GRAVITY \citep{2017A&A...602A..94G}, or MATISSE \citep{2022A&A...659A.192L}. The preset is sent to the VLTI Supervisor, which forwards it to the various subsystems of the VLTI: telescope, delay lines, STS, laboratory configuration, and GPAO. GPAO immediately initiates its (re)configuration and setup, in parallel to the telescope slewing. Relevant to GPAO, the preset contains the desired AO mode (see Sect.~\ref{sec:operational_modes}), the magnitude of the natural AO guide star, the coordinates of the AO guide star, the desired coordinate for the LGS guide star, and the on-axis coordinates of the telescope. The difference between these coordinates drives the position of the NXYT, the LXYT and of the Field Steering Mirror (FSM), and therefore the off-axis offset of the wavefront sensors at the Coud\'e focus. The NGS wavefront sensor is setup to a low amplification, and to a rate adapted to detect the given magnitude without risking a sensor over-illumination. A filter is possibly inserted to reduce the flux on very bright objects. If requested, the LGS wavefront sensor is setup at a fixed frame rate of \SI{1000}{\hertz} and an amplification gain of 400. The LGS system is also configured and the laser is propagated as soon as the telescope has reached the coordinate and follows a blind trajectory. The LSM of the Laser Launch Telescope is setup from a elevation-based pointing model, to compensate for the flexures, so that the laser spot immediately appears within the field-of-view of the LGS WFS. The LFOC device is setup to the expected distance of the Sodium layer at the time of the preset.

The acquisition with GPAO, fully automated by default, is composed of three steps: detecting the star (DETCGS), optimization of camera and loop parameters (OPTCGS), and closing the loop (STRTCAG). In LGS modes, these three steps are first applied to the laser guide star, and only then to the natural guide star. The DETCGS executes a raster with the QSM (for the NGS) or with the JM (for the LGS) and moves the telescope (for the NGS) or the LSM (for the LGS) to the position of maximum of flux. Then, the OPTCGS raises the amplification gain up to the maximum permissible value (42 in IR WFS or 1000 in LGS and VIS WFSs) and estimates the actual magnitude of the star. For the NGS VIS WFS only, this magnitude serves for an optimization of the camera amplification gain and frame rate. Apart from this specific point, the acquisition is strictly the same whether the natural guide wavefront sensor is IR or VIS. The DM/WFS registration is then optimized following the strategy presented in \citet{2024A&A...687A.157B}. It consists of modulating a subset of KL modes on the DM, and fitting the measured interaction matrix with a fast correlation method.  The last step of the optimization estimates the median flux per subaperture, and updates the threshold of the RTC below which the gradients are considered invalid and forced to zero. At this point, the STRTCAG is executed and all loops related to this WFS are closed, including offloads and secondary loops. Once all required wavefront sensor(s) are acquired, GPAO raises to state ``Coud\'e Guiding'', which informs the scientific instrument to pursue its own acquisition.

The automatic acquisition works well when there are no other bright targets within a few arcseconds around the guide star. For more complex fields, the operator has the possibility to manually execute the acquisition steps one by one to specifically select the correct guide star.

\subsubsection{Coordination with science observations}

A typical observation block at the VLTI lasts for one hour and includes several individual exposures, with a typical overhead of a few seconds between exposures. The GPAO OS is informed of the beginning and end of every exposure.

\paragraph{Telemetry} GPAO records \SI{30}{\second} of AO telemetry at the start of each science exposure. This telemetry includes all frames for the ``loop data'' (slopes, intensities, commands, chopping state) and a sub-sampling at \SI{10}{\hertz} for the ``pixel data'' (pixel stream). The files are stored on the observatory servers for one month, for technical use, but the ESO central archive system does not allow them to be made externally available. At the end of each exposure, the OS collects and forwards the necessary information to build the AO-related part of FITS header.

\paragraph{Passive Support} The Unit Telescopes incorporate a mechanism (M1PMOVE) to regularly re-center laterally the M1 mirror itself inside its cell. This is especially important when observing at low elevation. When triggered, the lateral motion of M1 creates a strong tip-tilt and a large step of optical path delay, which can make the fringe-tracker lose the tracking for several seconds. Therefore, the automatic, regular M1PMOVE mechanism is deactivated when the UTs are used for VLTI. Instead, GPAO requests the M1PMOVE only at preset.

\subsubsection{Maintenance and monitoring}

All calibrations are executed in day-time.
The calibration unit allows recalibrations to be performed of the internal alignment, pixel scale, reference slopes, and conversion gain and amplification gain of the camera, for the NGS VIS and the LGS wavefront sensors. No such tool exists for the NGS IR wavefront sensor. The dark level of all cameras are re-calibrated by closing the corresponding shutters. The PSIM parameters (see Sect.~\ref{sec:PSIM}) can be re-calibrated by measuring interaction matrices using the Nasmyth beacon. The DM reference vector used to flatten the DM in open-loop can be re-calibrated using the Nasmyth beacon. The template inverts a zonal interaction matrix keeping 1100 eigenvectors to create a zonal control matrix, then closes the loop for a few seconds without any modal or zonal leaks, and records the DM position.

The daily health check includes a measurement of the gain of all actuators of the DM, to verify their functionality. The mean gain is trended over time as well as the number of nonfunctioning actuators inside the clear aperture (currently only one known in GPAO2).  The daily health check also includes a measurement of the DM flat reference vector. It is used for trending only and the actual vector installed in the system is not updated automatically.

\subsection{Verifications}
\label{sec:performances}

\subsubsection{Transmission}
\label{sec:transmission}

Figure~\ref{fig:GPAO_NGS_VIS_flux_vs_mag} in Appendix~\ref{appendix:transmission} shows the measured flux as a function of source magnitude. From this the following zero points can be derived for GPAO 1 to 4 for a zero magnitude star in the Gaia \Grp{} band: $1.12,\;1.32,\;1.32,\;1.01\times10^{11}\,$e/s. The ratio between the best and the worst GPAO is 0.75. For the $40\times40$ WFS running at \SI{100}{\hertz}, the available flux is $\approx 5$ electrons per frame and per subaperture for a magnitude 13. This level of illumination corresponds to the operational limiting magnitude in NGS\_VIS mode. The zero-point of Gaia \Grp{} \citep[24.7619, see][]{2021A&A...649A...3R} corresponds to $5.45\times10^{11}$\,e/s when scaled to the collecting area of the Unit Telescopes (\SI{49.26}{\square\meter}) with respect to that of Gaia (\SI{0.723}{\square\meter}). The chromatic shape of GPAO matches well the one of the Gaia \Grp{} band-pass. The absolute peak transmission of Gaia \Grp{} being 74\%, the absolute peak transmission of the visible wavefront sensors of GPAO is in the range of 13.7\% to 17.9\% (end-to-end, all included).

In the infrared, a fit of Fig.~\ref{fig:GPAO_NGS_IR_flux_vs_mag} gives the following zero points for GPAO 1 to 4  for a zero magnitude star in the K-band: $1.62,\;2.06,\;1.51,\;1.42\times10^{10}\,$e/s. When running at \SI{100}{\hertz}, the illumination of five electrons per frame and per subaperture is reached at a magnitude 11, which again corresponds to the operational limiting magnitude of the NGS\_IR mode.

\begin{figure}[b]
\centering
\includegraphics[width=\columnwidth]{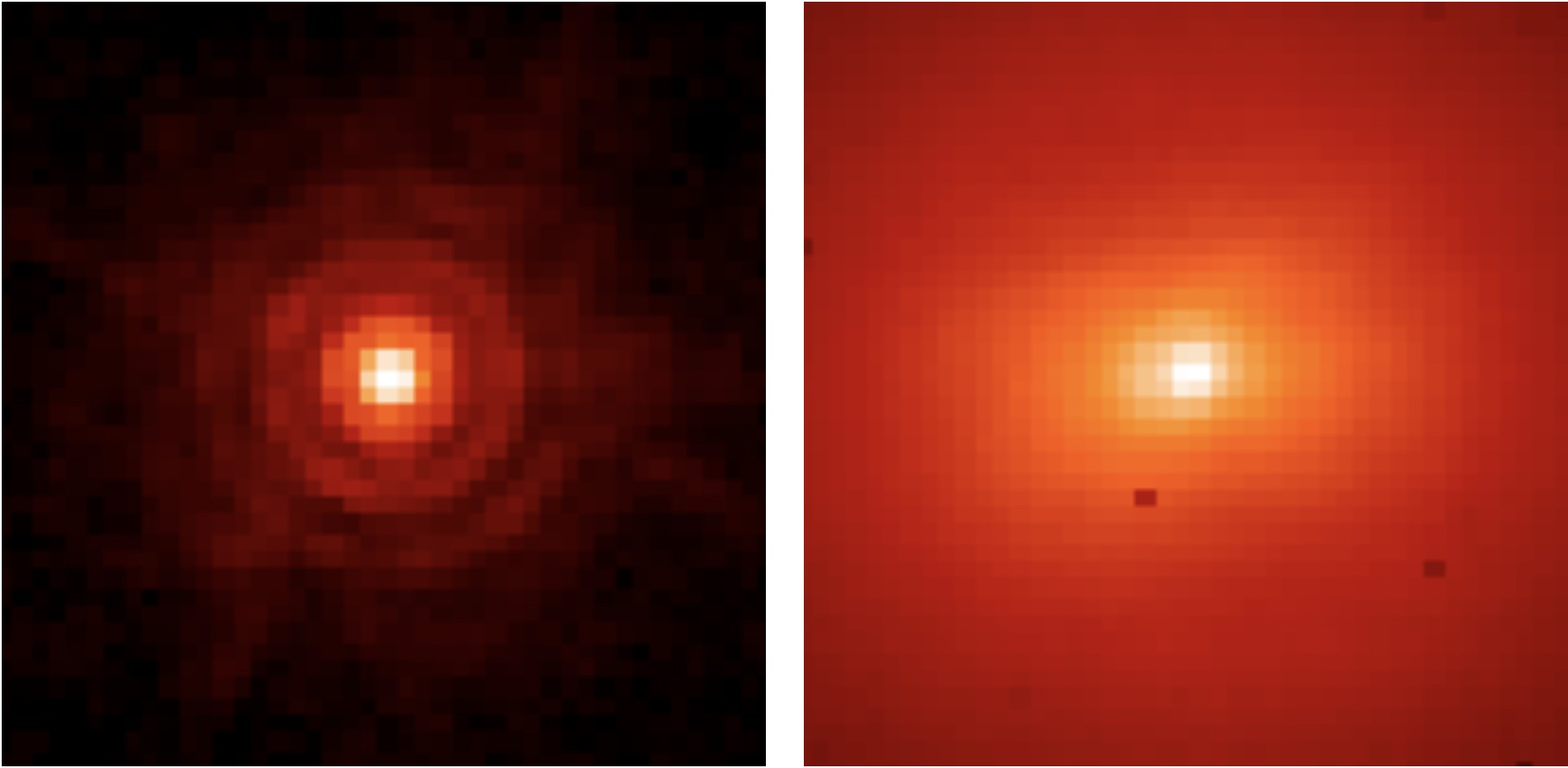}
\caption{Images of unresolved stars (point spread function) measured in K-band with the IRIS guiding camera of VLTI, corrected by GPAO in NGS\_VIS in the bright regime (left, $\Grp=5$, Strehl 77\%) and faint regime (right, $\Grp=12$, Strehl 11\%). The flux is represented in log scale.}
\label{fig:psf_image}
\end{figure}

\subsubsection{Transfer function}

Measurements and model of the transfer function is described in Appendix~\ref{app:transfer_function}. It includes the delay introduced by the integration time, the hold, and the mechanical response of the DM. It also includes an additional pure-delay term $T_d$ to account for the transferring, buffering and computations of data. We adjust this model to the measured rejection functions. Representative results are shown in Fig.~\ref{fig:GPAO_TF}. The rejection bandwidth at -3\,dB is at \SI{45}{\hertz} in the visible and \SI{50}{\hertz} in the infrared, while MACAO reached only \SI{18}{\hertz} \citep{2004SPIE.5490...47A}. The rejection bandwidth is now comparable to those of SPHERE or other extreme AO. 

The fitted integration gain $K_i$ are slightly smaller than the value actually used in the controller (0.5). There are several possible explanations such as optical gains or overestimated DM mechanical gains in our geometrical model. The slightly larger actual integration gain in the infrared sensor explains the higher rejection bandwidth. In theory the integration gains could be tuned closer to the stability limit to further improve the rejection. This was  discarded, however, because the optical gain can vary with the seeing conditions.

In the visible wavefront sensor, the expected pure-delay $T_d \approx \SI{830}{\micro\second}$ is the sum of the readout time of the OCAM2 (\SI{500}{\micro\second}, equal to the maximum frame rate), the time to extract the pixel from the camera (\SI{47}{\micro\second}, from manufacturer), the buffering in the frame-grabber (\SI{110}{\micro\second}, frame sent in 5 packets) the HRTC latency (\SI{80}{\micro\second}, measured at the oscilloscope), the time to transfer the command to the DM at 1\,GbE (\SI{21}{\micro\second}), and the buffering inside the DM electronic (\SI{70}{\micro\second}, from manufacturer). This prediction matches adequately the measured pure-delay of $T_d=\SI{856}{\micro\second}$, therefore validating our understanding of the dynamic of the system. A \component{total loop delay} is often computed by adding the pure-delay (above), half the DM rise time (\SI{400}{\micro\second}), and 1 full frame to account for the digitalized controller. This total loop delay is 2.4 frames at \SI{1}{\kilo\hertz} frame rate.

\begin{figure}[t]
\centering
\includegraphics[width=\columnwidth]{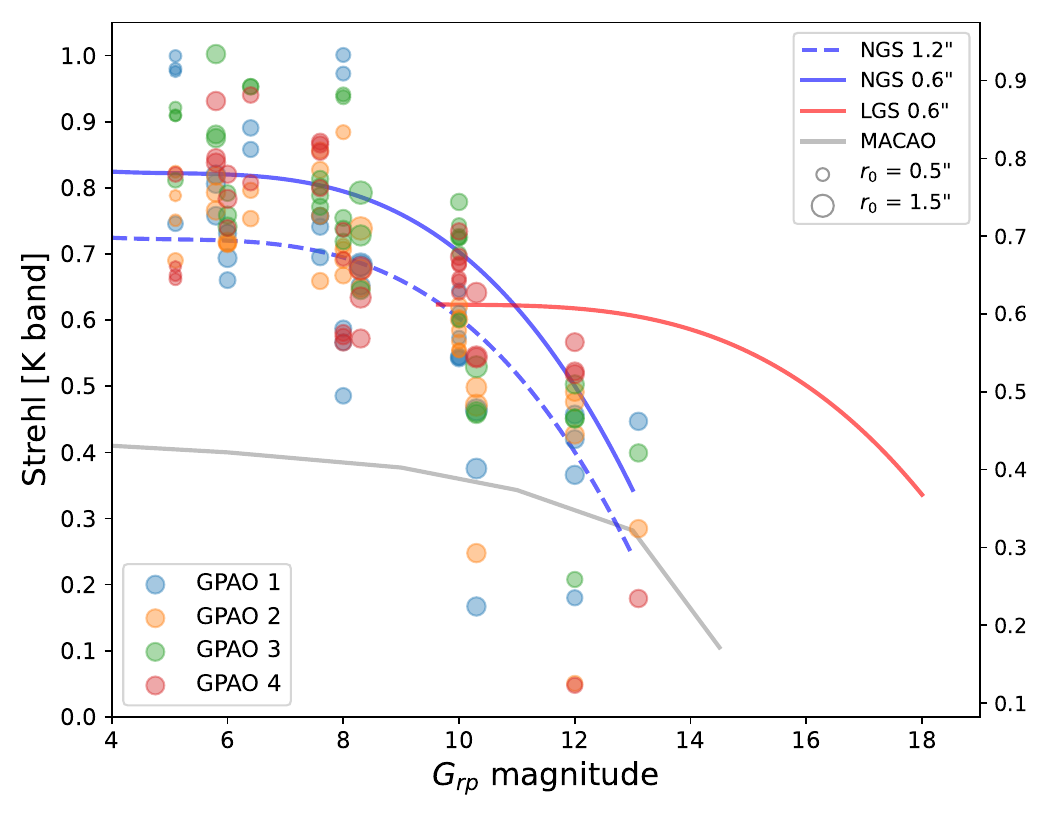}
\caption{Strehl in the K-band versus the Gaia \Grp{} magnitude of the guide star. Points indicate the measurements with GPAO in NGS\_VIS mode. The seeing conditions are proportional to the diameter of the point. The blue curves are the expectations for GPAO in  NGS\_VIS mode and two seeing values, the red curve is the expectation for GPAO in LGS\_VIS mode with seeing 0.6\arcsec{}. The gray curve shows the performance of the former MACAO system of VLTI.}
\label{fig:strehl}
\end{figure}

The NGS IR detector uses a reset and double-correlated read sequencer. It reduces the duty cycle, but also reduces the pure-delay with respect to our basic model of Appendix~\ref{app:transfer_function}. There is no buffering in a frame grabber. Finally, the computation time is shorter because there are fewer pixels and slopes to process. Altogether, it explains why the measured pure-delay ($T_d=\SI{515}{\micro\second}$) is shorter than in the visible wavefront sensor.

\begin{figure}[t!]
\centering
\includegraphics[width=\columnwidth]{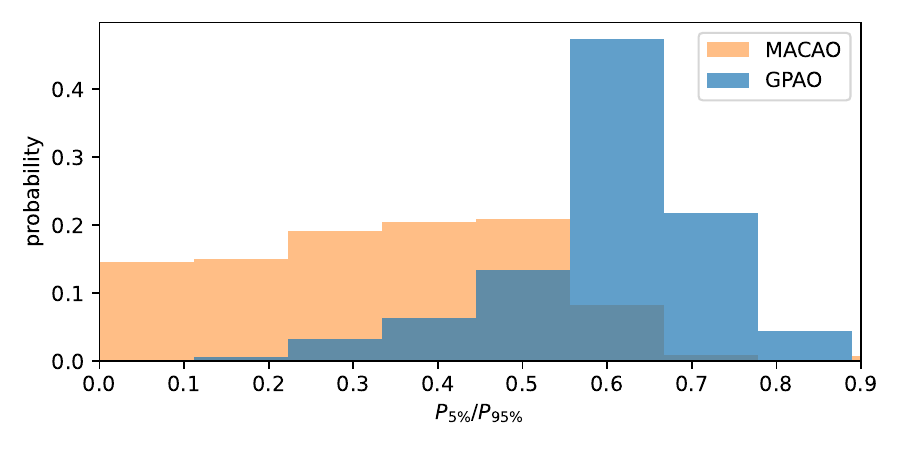}
\caption{Comparison between MACAO (entire year 2023) and GPAO NGS\_VIS (from December 2024 to May 2025) of the $P_{5\%}/P_{95\%}$ injection metric histograms for the GRAVITY fringe tracker. The flux dropouts, represented by $P_{5\%}/P_{95\%}$ values close to zero, are significantly reduced by GPAO.}
\label{fig:GPAO_MACAO_strehl_stability}
\end{figure}

\begin{figure}[b]
\centering
\includegraphics[width=\columnwidth]{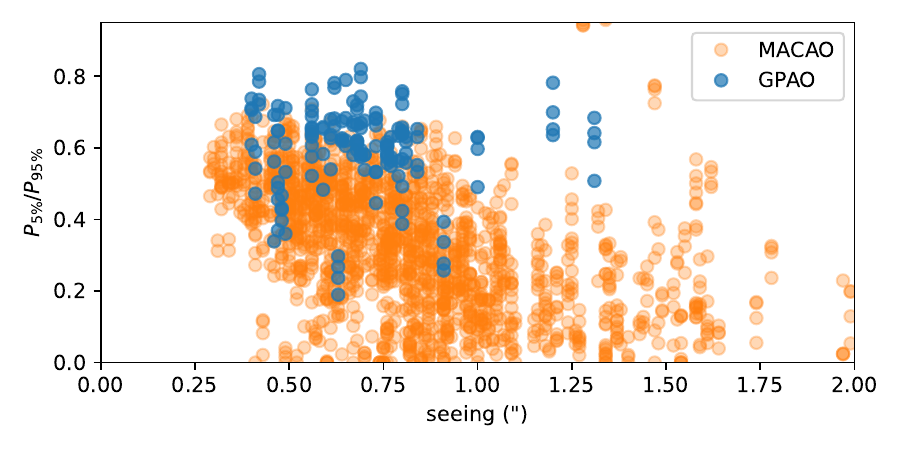}
\includegraphics[width=\columnwidth]{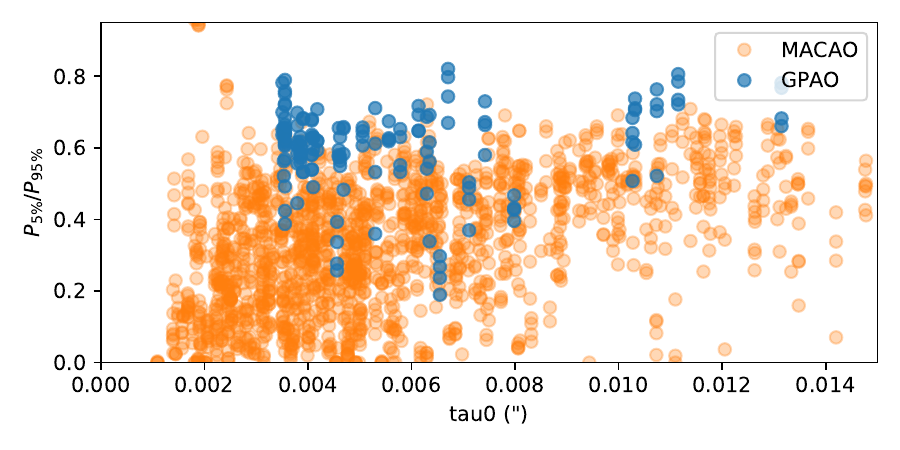}
\caption{Comparison between MACAO and GPAO of the $P_{5\%}/P_{95\%}$ injection metric histograms for the GRAVITY fringe tracker versus the atmospheric conditions.}
\label{fig:GPAO_MACAO_strehl_stability_versus_condition}
\end{figure}

\begin{figure*}[t!]
\centering
\includegraphics[width=0.9\textwidth]{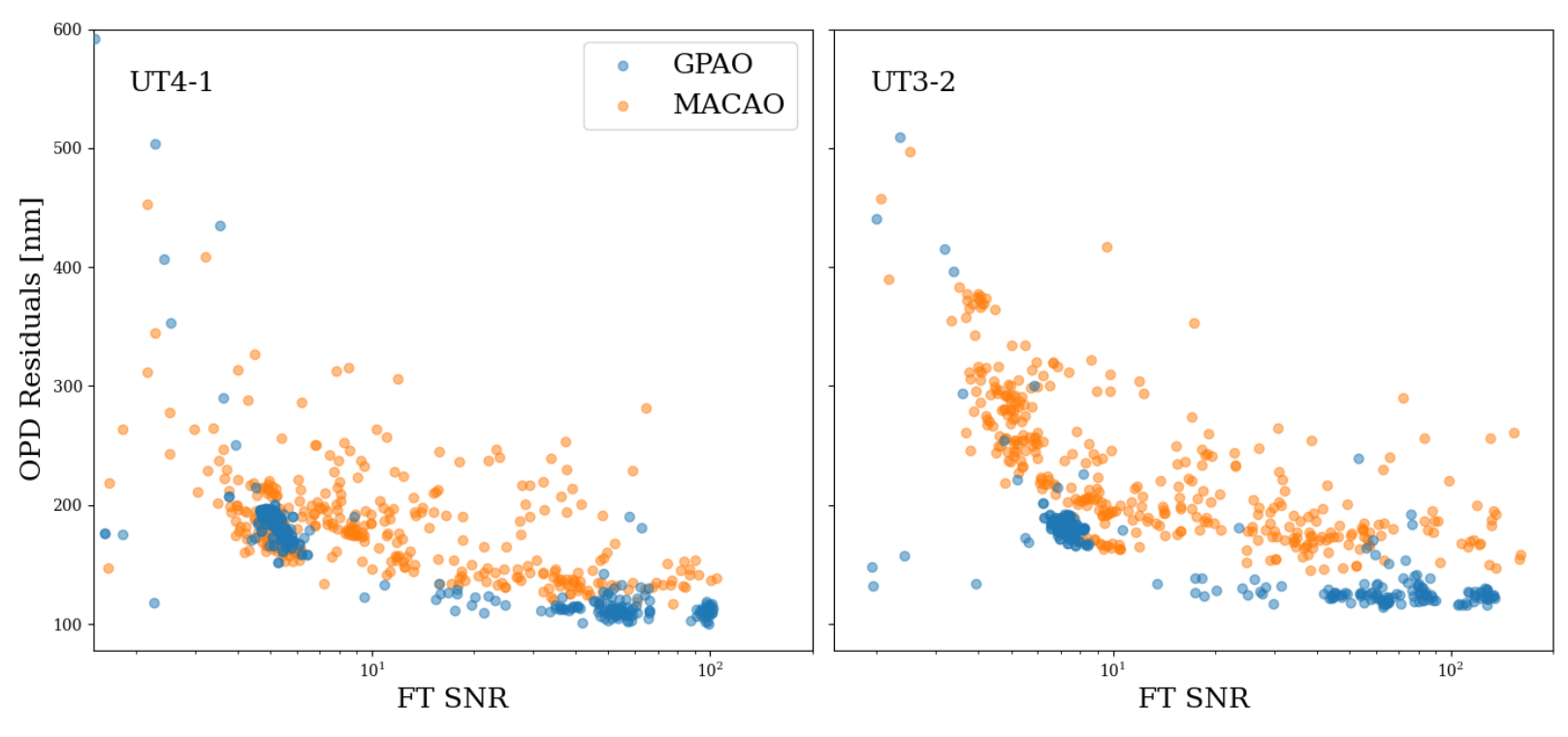}
\caption{Comparison between MACAO and GPAO of the optical path difference residual after fringe tracking, as a function of the signal to noise ratio in the GRAVITY Fringe Tracker. Data are shown for two independent baselines, which are representative of the performance of the entire array.}
\label{fig:GPAO_MACAO_OPD_residual}
\end{figure*}

\subsubsection{Piston control}
\label{sec:piston_control}

Following the method presented in \citet{2019A&A...629A..41W}, a piston conversion factor for the first controlled modes is measured by injecting a modulation on the DM and detecting it with the GRAVITY fringe tracker. The conversion factor is the ratio between the piston measured in \SI{}{\micro\meter}-rms and the modal modulation in \SI{}{\micro\meter}-rms. The modal basis of GPAO is defined to be piston-free when integrating the wavefront with a uniform weight over the telescope pupil. When accounting for the single-mode apodisation of the pupil by the GRAVITY fiber, the model predicts significant conversion factors for focus and the second spherically symmetric mode (and all modes with radial-only dependency). We measure a conversion factor of 0.57 for the focus (0.35 expected) and 0.07 for the first spherical (0.05 expected).

\subsubsection{Acquisition time budget}

The duration of the GPAO acquisition is measured from the moment the telescope delivers the beam (first cycle of telescope active optics), to the moment the GPAO adaptive optics loop is closed. Therefore, all setup activities that are parallelized with the telescope slew are not accounted for. The acquisition in NGS mode have a median duration of 110 seconds, with 95\% of the acquisitions lasting fewer than 180 seconds. The LGS acquisitions are typically 30 seconds longer than the NGS.

Changing GPAO mode requires some hardware (closing/opening shutters, moving HO/LO stage, turning filter wheels) to be reconfigured, but also the full shut down and restart of all the individual SRTC and HRTC processes. This action takes about 200 seconds.

\subsection{System-level performances}
\label{sec:system_performances}

\subsubsection{Strehl versus magnitude and seeing}
\label{sec:strehl}

The point spread function (PSF) of the system is commonly estimated by observing unresolved stars. Typical images of PSF observed with the IRIS guiding camera of VLTI are shown in Fig.~\ref{fig:psf_image}. The system was setup with the default NCPA, calibrated on the Nasmyth beacon with the method detailed in Sect.~\ref{sec:ncpa}. In the bright regime, the PSF is dominated by the diffraction pattern of the Unit Telescope aperture and spiders. In the faint regime, the PSF is dominated by the residual of uncorrected turbulence. The elongated aspect of the halo indicates the direction of the wind because the system is dominated by the temporal error. Most of the time, the PSF is dominated either by the diffraction of the aperture or by the residuals of the fast atmospheric turbulence, as presented here. Occasionally, the PSF is dominated by the so called ``low wind effect'' as also seen in others high contrast instruments \citep[see for instance][]{2018SPIE10703E..2AM}. Preliminary analysis indicates that the effect is more pronounced in telescopes UT1-2, which matches the fact that the spiders of UT3-4 have been re-coated with special painting to increase their thermal coupling with surrounding air. Re-coating the spiders of UT1-2 will be implemented in the next year as part of the GRAVITY+ project.

Figure~\ref{fig:strehl} presents the measurements of Strehl collected during the commissioning nights versus seeing and magnitude of the guide star for the NGS\_VIS mode. Important to notice, the system was still under test and optimization at the time. Especially, the NCPA were not always properly calibrated. Nevertheless, the values closely match the predictions. The peak performances are significantly improved with respect to the former MACAO that was reaching a top Strehl of 0.4 on the brightest targets in the best atmospheric conditions. The limiting magnitude of GPAO is about the same as MACAO. This is expected since the faint-end will be covered by the LGS mode once available.

\begin{table}[b]
\caption{Performances of the four modes of GPAO and of the former MACAO and CIAO systems.}
\label{tab:performance_summary} 
\centering
\begin{tabular}{l c c c c}
\hline\hline
   \\[-2.2ex]
    AO & Strehl\tablefootmark{a} & Bright & Faint\tablefootmark{b} & Filter \\
    &  K-band & [mag] & [mag] & \\
   \hline\\[-2.1ex]
   MACAO  & 40\,\% & 9 & 13 & V \\
   NGS\_VIS & 75\,\% & 8 & 12 & \Grp{}\\
   LGS\_VIS\tablefootmark{c} & 60\,\% & 13 & 17 & \Grp{} \\
   \hline\\[-2.1ex]
   CIAO  & 40\,\% & 9 & 12 & K\\
   NGS\_IR & 40\,\% & 9 & 12 & K \\
   LGS\_IR\tablefootmark{c} & 60\,\% & 9 & 12 & K
   \\ \hline
\end{tabular}
\tablefoot{
\tablefoottext{a}{Strehl obtained under nominal weather conditions on a bright target.}
\tablefoottext{b}{Operational limit with reduced performances.}
\tablefoottext{c}{The performances of the LGS modes are preliminary and have been only roughly verified on sky .}}
\end{table}

\subsubsection{Injection stability}
\label{sec:injection_stability}

Interferometric fringe tracking is particularly sensitive to occasional flux dropouts, which can make the fringe tracker lose the white-light position and lead to a long recovery time before a truly coherent integration can resume. \citet{2010A&A...524A..65T} theorized the effect of improved AO in the coupling into a single mode fiber, with a specific emphasis on the stability of the coupling for optical interferometry. We quantified the improvement brought by GPAO by using the same metric as used for NAOMI in \citet{2019A&A...629A..41W}. This metric, noted $P_{5\%}/P_{95\%}$, represents the ratio of the flux at low 5\% injection to the flux at high 95\% injection. It tends to one for a perfectly stable injection, and tends toward zero in the presence of regular flux dropouts. Figure~\ref{fig:GPAO_MACAO_strehl_stability} compares the histogram of the $P_{5\%}/P_{95\%}$ metric for GPAO (blue) and MACAO (orange) for each of the UTs. The measurements are from the GRAVITY Fringe Tracking instrument, in the K-band. The clear shift toward higher $P_{5\%}/P_{95\%}$, with a median of 0.31 for MACAO and 0.62 for GPAO, demonstrates a tremendous improvement in the stability of the flux injected into the single mode fiber of the fringe tracker.

A decisive advantage of GPAO is its improved robustness against degraded atmospheric conditions, illustrated in Fig.~\ref{fig:GPAO_MACAO_strehl_stability_versus_condition}. The former MACAO system was saturating for seeing larger than 1\arcsec{}, with devastating effects on the Strehl because of the extended influence functions of the curvature deformable mirror. Thanks to the improved stroke, and the use of a mirror technology with localized influence functions, GPAO safely operates even with seeing larger than 1.5\arcsec{}. GPAO can also operate in faster coherence time thanks to a maximum frame rate of \SI{1000}{\hertz} with respect to \SI{500}{\hertz} for MACAO. When comparing the statistic of turbulence at Paranal, it corresponds to an increase in operational time for VLTI by 30\% (now reaching nearly all conditions).

\subsubsection{Fringe tracking performances}
\label{sec:fringe_tracking_performances}

Figure~\ref{fig:GPAO_MACAO_OPD_residual} compares the optical path delay residuals as a function of the S/N of the Fringe Tracker for GPAO NGS\_VIS and MACAO, for two independent baselines. The faint end displays the expected dependency between the S/N and the performance for a noise-limited system. The bright end shows the expected plateau of a rejection-limited system. The improvement brought by GPAO is obvious, with residuals just above \SI{100}{\nano\meter} compared to \SI{150}{\nano\meter} when the fringe tracker was fed by MACAO. The much smaller scatter is synonymous of more reliable operation.

This positive outcome of the installation of GPAO was not anticipated. The best explanation comes from the change in deformable mirror technology. The MACAO curvature deformable mirror, with influence functions highly coupled to the piston mode, was replaced by the ALPAO deformable mirror with localized influence functions. We believe that the MACAO system was somehow coupling the high spatial and temporal frequency atmospheric perturbations back into a high temporal frequency piston.

\subsection{Discussion on bright and faint end}
\label{sec:discussion_bright_faint_end}

Table~\ref{tab:performance_summary} gives a condensed overview of the GPAO performances.
On the one hand, the GPAO system in GPAO\_VIS mode outperforms the former MACAO system on bright stars by a factor larger than two. This is expected because MACAO was designed as a general purpose facility while GPAO in NGS\_VIS mode is optimized for the peak performance in the bright regime.  On the other end, the faint end of GPAO NGS\_VIS is similar to MACAO, when considering the mean Strehl only (see Sect.~\ref{sec:strehl}). However, combined with the improvement in Strehl stability (see Sect.~\ref{sec:injection_stability}) and the improvements in OPD residuals (see Sect.~\ref{sec:fringe_tracking_performances}), the interferometric performances of the VLTI with GPAO in NGS\_VIS mode are significantly increased with respect to MACAO, even in the faint end. This is demonstrated by several observations both for the bright end and the faint end as presented in the next section, that were impossible with MACAO.

The limiting magnitude of GPAO in NGS mode is close to its fundamental limit of $\Grp\approx13.1$ (see Appendix~\ref{appendix:fundamental_limits}). At first order, the difference matches the transmission detailed in Sec.~\ref{sec:transmission}. The limiting magnitude of the LGS mode is significantly further away from its fundamental limit of $\Grp\approx20.3$. It comes from the choice of using $4\times4$ subapertures in the low-order sensor instead of the full telescope aperture.

\begin{figure*}[t]
    \centering
    \includegraphics[width=0.85\textwidth]{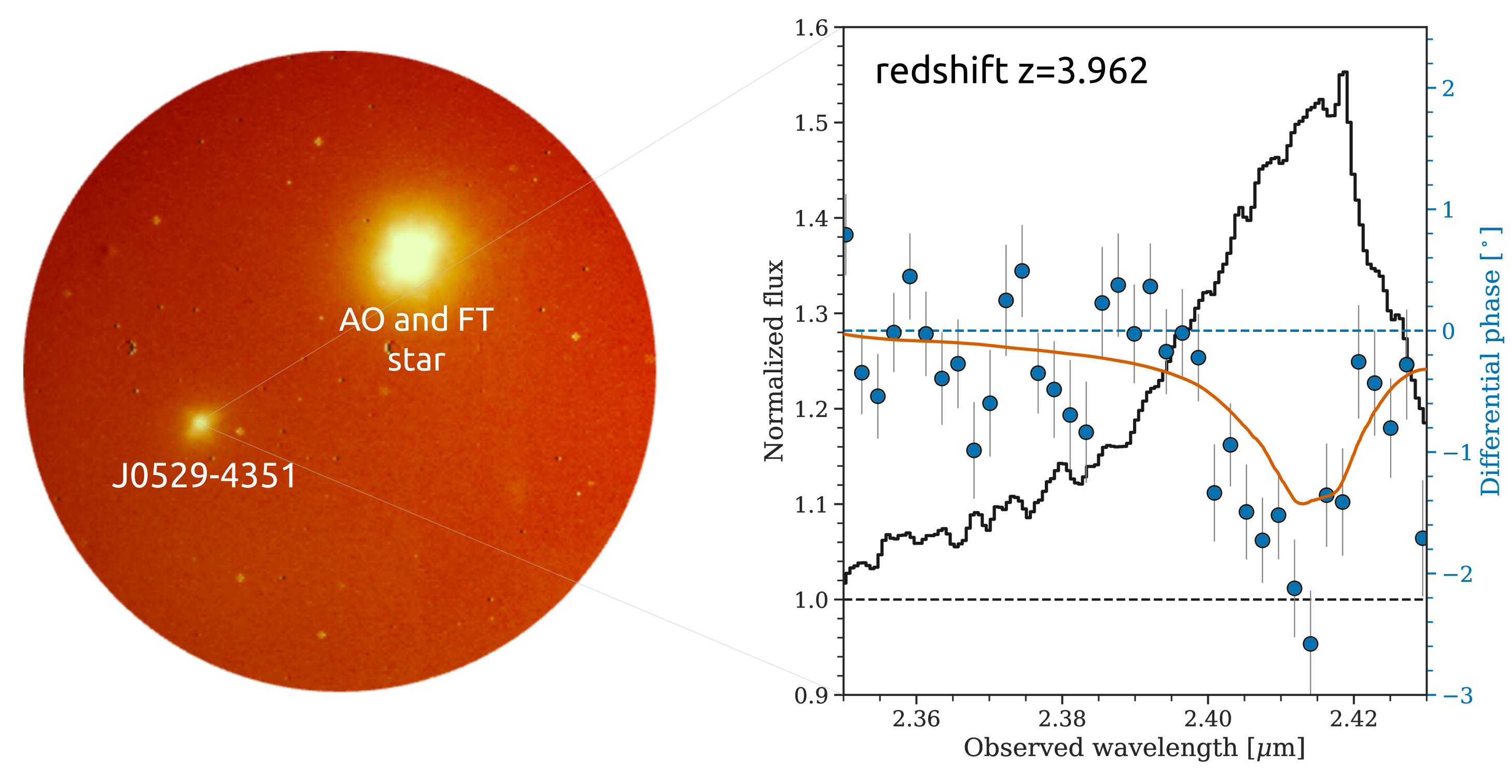}
    \caption{\textbf{Left:} GRAVITY H~band Acquisition Camera image of the observation of J0529-4351 showing the quasar and AO/FT reference star. On sky the separation is 22\,\arcsec{} but the VLTI optics bring them to GRAVITY in separate beams with a separation of about  2\,\arcsec{} \textbf{Right:} Averaged differential phase spectrum of J0529-4351 (blue points) across the H$\beta$ emission line (black line). An initial BLR model fit to the line profile and differential phase (red line) shows a much smaller BLR and SMBH mass.}
    \label{fig:SCIENCE_qso}
\end{figure*}

\section{Enabling new science}
\label{sec:science}

This section presents the new science enabled by GPAO coupled to on-sky interferometric observations. We illustrate this new range of capabilities with several archetypical objects that were acquired during the end-2024 commissioning.

\subsection{High-redshift quasar z=4}

Extragalactic science depends critically on sensitivity. The implementation of G-Wide (\citet{GWide2022}, see Sect.~\ref{subsec:gravityplus}) enabled the first dynamical mass measurement of a $z\sim2$ Super Massive Black Hole (SMBH) \citep{GCollab2024_z2}. To push even further into the early Universe and to build statistical samples probing SMBH-galaxy coevolution across cosmic time requires an increase in both sensitivity and sky coverage for GRAVITY, because of the need for off-axis reference stars to perform AO correction and fringe tracking. As discussed in Sect.~\ref{sec:system_performances}, the new GPAO systems with high-order corrections achieve both through the improved Strehl and fiber injection efficiency that increases the S/N on the interferometric observables and pushes the fringe tracking limiting magnitude to fainter objects. 

Expanding the capabilities of GRAVITY to dynamically measure black-hole masses at earlier times and fainter objects is also very timely given recent results from the James Webb Space Telescope (JWST). JWST-discovered broad-line AGNs at $z>4$ suggest a fast growth of black holes that outpaces the growth of their host galaxy and leads to over-massive black holes \citep[e.g.][]{Maiolino_jwst_agn}. These masses are derived from local scaling relations extrapolated to high redshift and indeed significant changes in the gas structure can lead to up to two orders of magnitude difference in the inferred mass \citep[e.g.][]{Maiolino_xray_weak,Naidu2025}. Direct measurements of SMBH masses are therefore critical in our understanding of black-hole growth.

Here, we illustrate the observations of a $z\sim 4$ quasar with GPAO-NGS in combination with the G-Wide mode. We selected \object{SMSS J052915.80-435152.0} ($\Kmag=13.9$), hereafter called J0529-4351, which was discovered to be ``the most luminous quasar in the Universe'' \citep{Wolf2024}. Its high luminosity would predict a very large BLR differential phase signal 
and its redshift places both the H$\beta$ and H$\gamma$ recombination lines in the $K$-band. J0529-4351 is also located 22 arcsec away from a star with $\Grp=12.1$ and $\Kmag=11.0$ (see Fig.~\ref{fig:SCIENCE_qso} left), making it a good test of the capabilities of GPAO-NGS at the faint end as well as the new FT limiting magnitude. 

\begin{figure}[ht!]
\centering
\includegraphics[width=0.9\columnwidth]{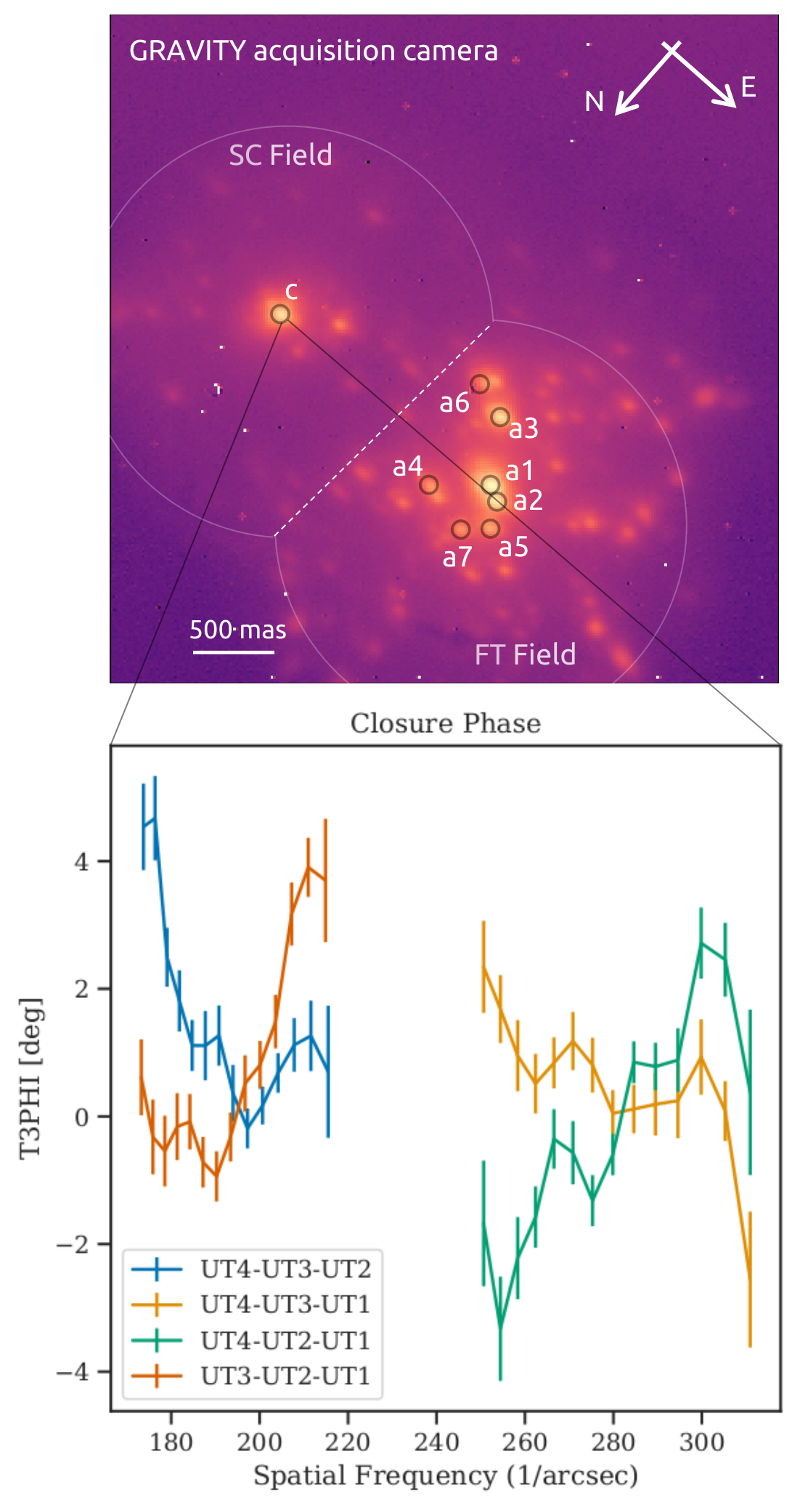}
\caption{\textbf{Upper panel:} RMC\,136 seen on GRAVITY Acquisition Camera. \textbf{Lower panel:} Closure-Phases of RMC\,136c observed with G-Wide and GPAO. These observations are close to the limiting magnitude in NGS\_VIS mode and will greatly benefit from the LGS next year.}
\label{fig:rmc136}
\end{figure}

We observed J0529-4351 with the MEDIUM spectral resolution grism and COMBINED polarization mode over five nights. The atmospheric conditions ranged from 0.4\arcsec{} to 1.1\arcsec{}, and required overall good atmosphere conditions given the large separation between the AO/FT star and the quasar, which affects the anisopistonic errors. The total usable data is about \SI{7.8}{\hour} on-target, for which the improved performance of the AO for seeing $\geq 0.5\arcsec{}$ is crucial, and previously not possible with MACAO. The raw data with DIT=\SI{100}{\second} and NDIT=4 were reduced through the GRAVITY pipeline to produce complex visibilities and differential phases. The differential phase data were analyzed following the methodology described in \cite{GCollab2024_z2}. In the right panel of Fig.~\ref{fig:SCIENCE_qso}, we show the total photometric flux of J0529-4351 across the H$\beta$ emission line as well as the average differential phase spectrum by combining the longest three baselines (UT1-UT4, UT1-UT3, and UT2-UT4). The redshift of the quasar places the line near the edge of the wavelength range of GRAVITY and therefore cuts off part of the red wing. Nevertheless, we strongly detect the H$\beta$ line that peaks at \SI{2.41}{\micro\meter} and has a FWHM of $\sim\SI{3200}{\kilo\meter\per\second}$ with an underlying blue wing offset by $\sim\SI{2000}{\kilo\meter\per\second}$ and a FWHM of $\sim\SI{7000}{\kilo\meter\per\second}$. We measure a typical noise of $\sim 1^\circ - 2^\circ$ per spectral channel and per baseline within the H$\beta$ line for the total observing time.
The observations show a strong dip in differential phase, which demonstrates that we spatially resolve the BLR in J0529-4351. However, the differential phase signal does not follow a canonical ``S-shape'', which indicates a motion departing from pure-Keplerian motion. This is a potential signature of significant radial motion in J0529-4351, which has also been reported in GRAVITY observations of local AGNs \cite{GRAVITYCollab2024_agn}, but which has never been observed at high-redshift. 
The modeling of the BLR in this complex environment would indicate a BLR size and SMBH mass a factor of two smaller than the size predicted from scaling relations, with a measured size from GRAVITY of \SI{1.2}{pc}, and a SMBH mass of $10^{9.6}\,\Msun$. We caution the reader these are only preliminary results and a full analysis of the GRAVITY data for this quasar, including the H$\gamma$ line, is beyond the scope of this paper and will be presented in an upcoming dedicated paper.

\subsection{Extragalactic stellar interferometry}

Stellar physics has been a dominant field of optical/infrared interferometry (or ``stellar interferometry''). With the capability to observe faint targets and wide off-axis fringe-tracking targets, it becomes possible to extend stellar interferometry outside of the Milky Way in our local Universe. We demonstrate this capability on the Large Magellanic Cloud (LMC), which was observed on 27 September 2024 using GPAO and G-Wide. We focused on the dense central star cluster R136 of the Tarantula Nebula that hosts a large concentration of O and Wolf-Rayet stars within a 5 parsec region \citep{Feast1960}, and are the most massive stars known currently in the Universe \citep{Crowther2010}. These observations allowed the acquisition of targets in crowded fields to be validated, using the WFS technical camera for the acquisition of the AO star (see ~\ref{sec:wfs}), and the GRAVITY Acquisition Camera in G-Wide mode \citep{GWide2022}, see Fig. \ref{fig:rmc136}, upper panel. We observed the star R136c (\object{BAT99 112}) with a magnitude $\Kmag=11.31$, using the star R136a1 at a separation of $3.35\arcsec{}$ for AO ($\Vmag=12.8$) and FT ($\Kmag=11.15$). The observations were performed in good weather conditions, with seeing $\sim 0.45\arcsec{}$ and wind speed $\sim \SI{7}{\meter\per\second}$. In only 8 min of exposure on source (DIT=\SI{3}{\second}) and a spectral resolution $R\simeq20$, the precision obtained on the closure phases is of the order of $\sim 1 \deg$ per channel (see Fig. ~\ref{fig:rmc136}, lower panel). We analyzed the closure phases to look for a signature of binarity. Despite the short observing time on target, we can rule out binary companions in R136c with a flux ratio of $\geq 0.1$ and separation $\gtrsim \SI{5}{\milli as}$. These observations exclude massive companions ($M \gtrsim 20 \,\Msun$) at large separation $\gtrsim \SI{250}{au}$. Significantly stronger constraints on the flux ratio and the separation $100-\SI{1000}{au}$ can be obtained in the future with longer integration and improved uv-coverage. These constraints are very complementary to RV data, which are mainly probing periods shorter than tens of years, corresponding to companions with a separation $\lesssim \SI{100}{au}$.

\begin{figure}[t]
\centering
\includegraphics[width=0.9\columnwidth]{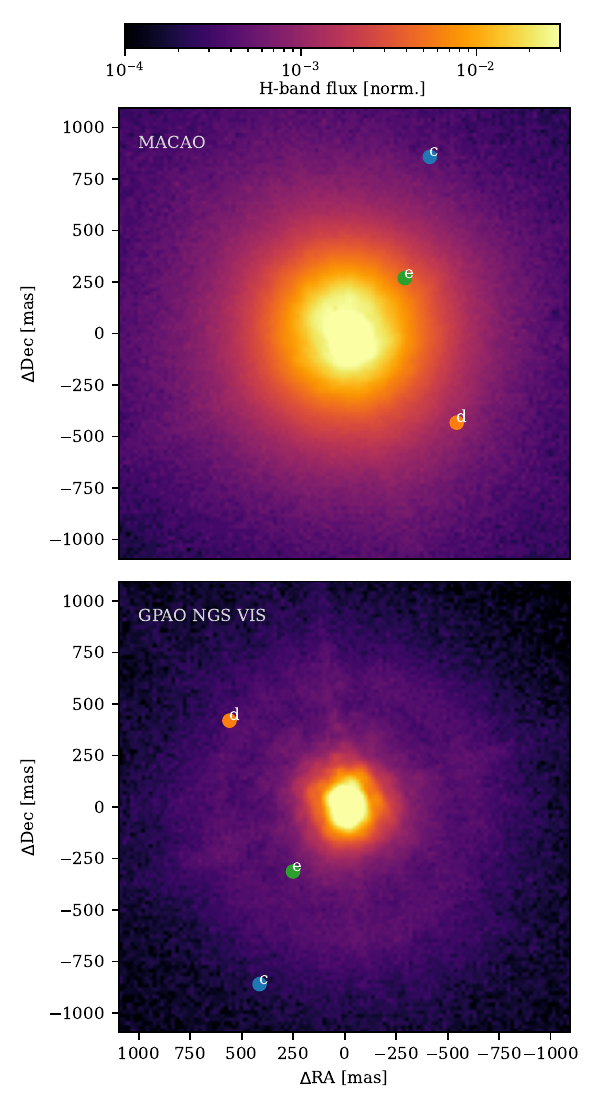}
\caption{Image of the GRAVITY Acquisition Camera (H-band) showing the saturated PSF around HR\,8799, when observed with MACAO (top) and with GPAO in NGS\_VIS (bottom). The scale is normalized to the maximum of the image. The ring structure corresponds to the spatial cutoff frequency of GPAO when controlling 500 modes. The positions of the known exoplanets are shown for reference.}
\label{fig:SCIENCE_HR8799_acqcam}
\end{figure}

\begin{figure}[t!]
    \centering
    \includegraphics[width=\columnwidth]{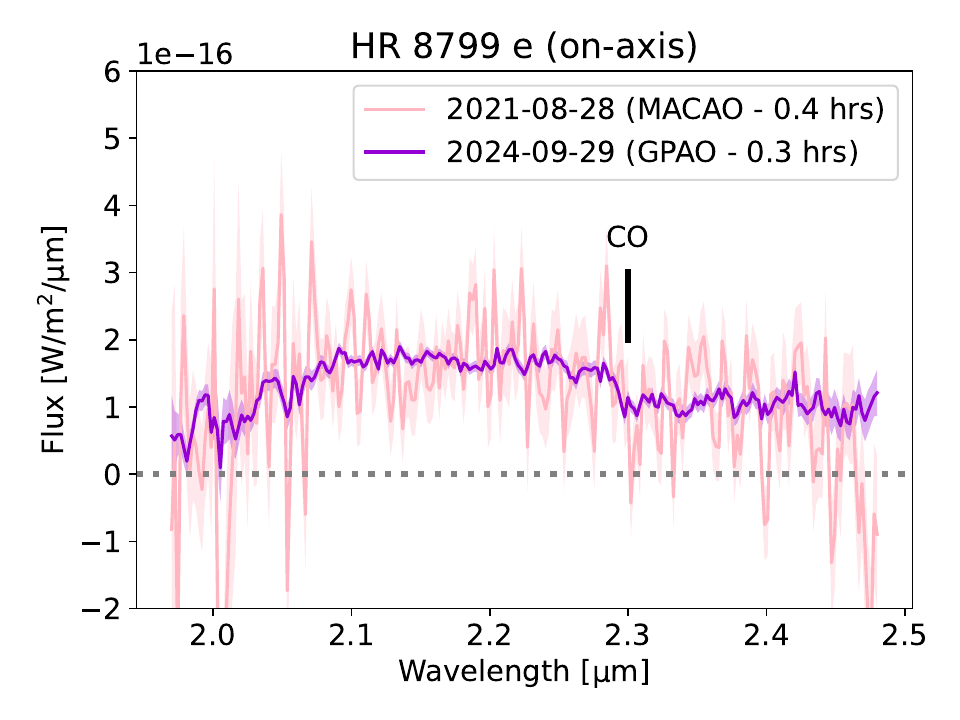}
    \caption{GRAVITY on-axis spectrum of the young giant planet HR~8799~e obtained with the new GPAO system and with the old MACAO system.}
    \label{fig:hr8799e}
\end{figure}

\subsection{High-resolution spectroscopy of exoplanets}

The upgrade from MACAO to the high-order correction of GPAO is crucial for the observations of exoplanets. This science case benefits both from the improved rejection of the starlight and the better flux injection of the planet in the science fiber. Figure~\ref{fig:SCIENCE_HR8799_acqcam} showcases the reduction in the glow around the bright star HR 8799. The brightness of the halo is reduced by a factor $9.1$, $2.8$ and $2.0$ at the separation of the planets e (\SI{400}{\milli as}), d (\SI{690}{\milli as}) and c (\SI{955}{\milli as}) respectively. In the following, we illustrate the resulting improvement of scientific capabilities with high-resolution spectroscopy of exoplanets and brown dwarfs. 

\subsubsection{High-resolution spectroscopy of HR~8799~e}

The young giant planet \object{HR 8799 e} was the first exoplanet detected using optical interferometry \citep{GCollab2019_hr8799}. This object is located at a separation $\sim0.4\arcsec{}$ from its host star and has a magnitude of 15.9~mag in the K band. HR\,8799\,e was re-observed with GRAVITY dual-field using GPAO instead of MACAO during the night of 28 September 2024, with a seeing $\sim 0.7\arcsec{}$. The observing time on target was only \SI{0.3}{\hour}, but shows the significant improvement in terms of sensitivity compared to previous observations.
The new spectrum obtained with GPAO is shown in Fig.~\ref{fig:hr8799e}, together with an old HR\,8799\,e spectrum from 28 August 2021 obtained with MACAO in \SI{0.4}{\hour} on HR\,8799\,e, both with a spectral resolution $R\sim 500$. 
For a comparable exposure time on target, an improvement in the S/N of a factor of $\sim 10$ can be clearly seen. This improvement is both consistent with an improvement of flux injection in the SC fiber by a factor $\sim 3.5$ and the reduction of noise by a factor $\sim 2.5$. The improvement of flux injection is consistent with the increase in the Strehl ratio for a $\Grp=5.4$ star from GPAO to MACAO, as shown in Fig.~\ref{fig:strehl}. The noise reduction originates from a better rejection of stellar photons and longer integration (DIT=\SI{100}{\second} compared to DIT=\SI{30}{\second} in 2021), the latter being due to improved fringe-tracking and reduced vibrations, as shown in Section \ref{sec:fringe_tracking_performances}. In only \SI{0.3}{\hour}, it is possible to clearly identify the CO ro-vibrational lines at \SI{2.3}{\micro\meter} in the spectrum of HR\,8799\,e. With long exposure, this new capability opens up the possibility of high signal-to-noise ratio spectroscopy in order to measure accurate C/O elemental abundances or time-resolved spectroscopy for young exoplanets, such as HR\,8799\,e or $\beta$\,Pic\,b.

\subsubsection{Spectroscopy down to \texorpdfstring{$\Kmag\sim21$}{Kmag sim 21}}

\begin{figure}[t]
    \centering
    \includegraphics[width=\columnwidth]{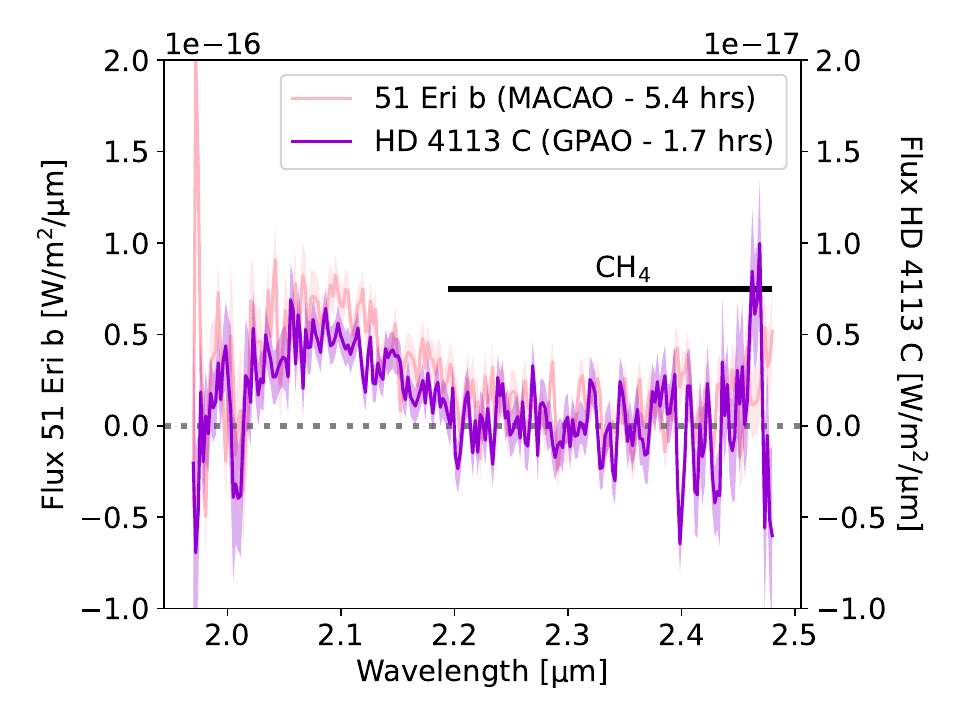}
    \caption{GRAVITY off-axis spectrum of the brown dwarf companion HD\,4113\,C obtained with the new GPAO system compared to the off-axis spectrum of the exoplanet 51\,Eri\,b obtained with the old MACAO system. We note the two different y-axes for the two objects, which show that a spectrum with a similar S/N can now be obtained for an approximately ten times fainter object.}
    \label{fig:hd4113c}
\end{figure}

We observed the ultra-cool brown dwarf \object{HD 4113 C} \citep{Cheetham2018} with GRAVITY and GPAO during the November 2024 commissioning run, together with the FAINT mode on GRAVITY \citep{Widmann2022}. This substellar companion is located at a separation of 0.7\arcsec{} from its host star, with a magnitude about 21.0~mag in the K band (Vega magnitude), significantly fainter than the faintest companion successfully observed with the old MACAO system, which was 51\,Eri\,b at about 18.5~mag in the K band at a similar separation. The observations were obtained in average weather conditions with seeing $\sim 0.6-1\arcsec{}$. The exposure on HD\,4113\,C were followed by a swap on a binary calibrator HD\,25535 \citep{Nowak2024} and reduced using the standard exoGravity tools\footnote{https://gitlab.obspm.fr/mnowak/exogravity} \citep{GCollab2020_betapic}. We used long exposure DIT=\SI{100}{\second} to maximize the S/N. In about \SI{1.7}{\hour} integration time on-target, we obtain an astrometry accuracy of about $\sim 50-\SI{100}{\micro as}$. We measure the spectrum of HD\,4113\,C with $R\sim 500$, showing a strong methane absorption band from \SI{2.1}{\micro \meter} to \SI{2.45}{\micro \meter}. The typical sensitivity obtained of HD\,4113\,C in \SI{1.7}{\hour} observations is comparable to the sensitivity obtained on 51\,Eri in \SI{5.4}{\hour} observations, as shown in Fig. \ref{fig:hd4113c}, which illustrates the major improvement in the S/N between MACAO and GPAO. From previous observations, HD\,4113\,C has an estimated effective temperature $T\sim500-\SI{600}{\kelvin}$ and a surface gravity $\mathrm{log} \, g=5$ \citep{Cheetham2018}. Its effective temperature is in tension with the isochronal mass estimate of $\sim 36\,\mathrm{M_{jup}}$, making this target a potential binary brown dwarf candidate, as recently demonstrated in Gl229\,B using GRAVITY resolving power \citep{Xuan2024}.

\begin{figure}[b]
\centering
\includegraphics[width=0.5\textwidth]{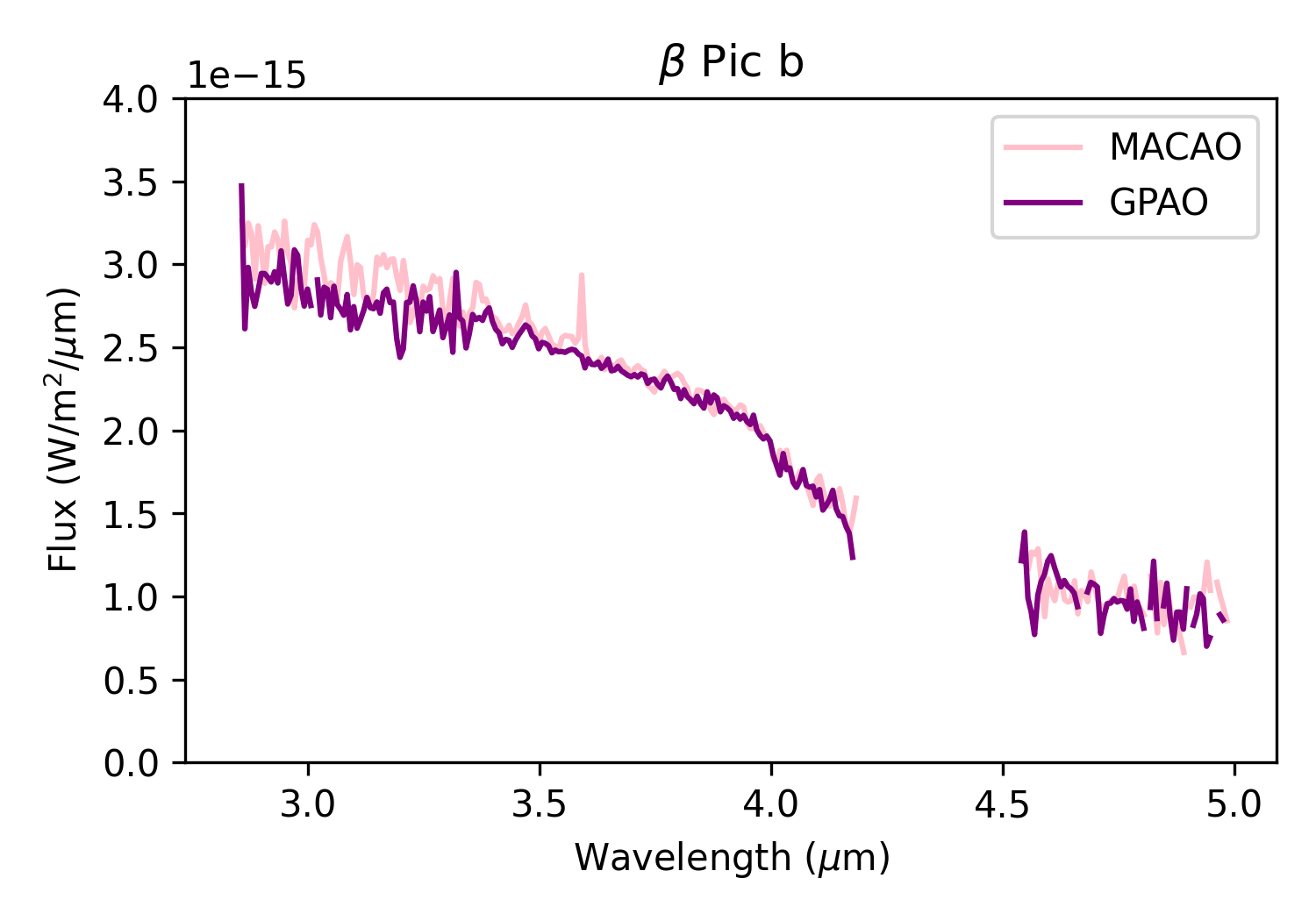}
\caption{Spectrum of $\beta$\,Pic\,b in the L and M bands obtained with MATISSE using MACAO and GPAO scaled by a factor of 4/3.}
\label{fig:betaPic_b_MATISSE}
\end{figure}

\subsubsection{Mid-infrared L+M spectroscopy}

The planet $\beta$\,Pictoris\,b (\object{HD 39060 b}) was observed with GPAO and MATISSE, using the GRA4MAT mode \citep{Woillez2024}. The observations were carried out during the night of 17 November 2024, with typical seeing conditions of $\sim 0.6\arcsec{}$. The MATISSE observations were obtained in L/M bands \citep{2022A&A...659A.192L}, using the MEDIUM spectral resolution $R\sim500$. From the coherent flux, the spectrum of $\beta$Pic\,b was extracted, following the methodology described in \cite{Houlle2025}. Figure~\ref{fig:betaPic_b_MATISSE} shows a comparison of the new spectrum obtained with the help of GPAO with that obtained with the help of MACAO extracted in the same way.  We find that the MATISSE/MACAO spectrum is scaled by a factor $\approx4/3$ ($+30\%$) compared to the MATISSE/GPAO one. This means that, in the MACAO observation, either the flux of the star is underestimated or the flux of the planet is overestimated; or vice-versa in the GPAO observation. MATISSE uses spatial filtering, and the GRA4MAT narrow off-axis mode was still in development at the moment of the MACAO observation, with visible drifts in the data. Nevertheless, the comparison of both spectra show the GPAO one has a 20\% higher S/N  compared to the MACAO one in the same amount of time spent on source.

\begin{figure}[t]
    \centering
    \includegraphics[width=0.87\columnwidth]{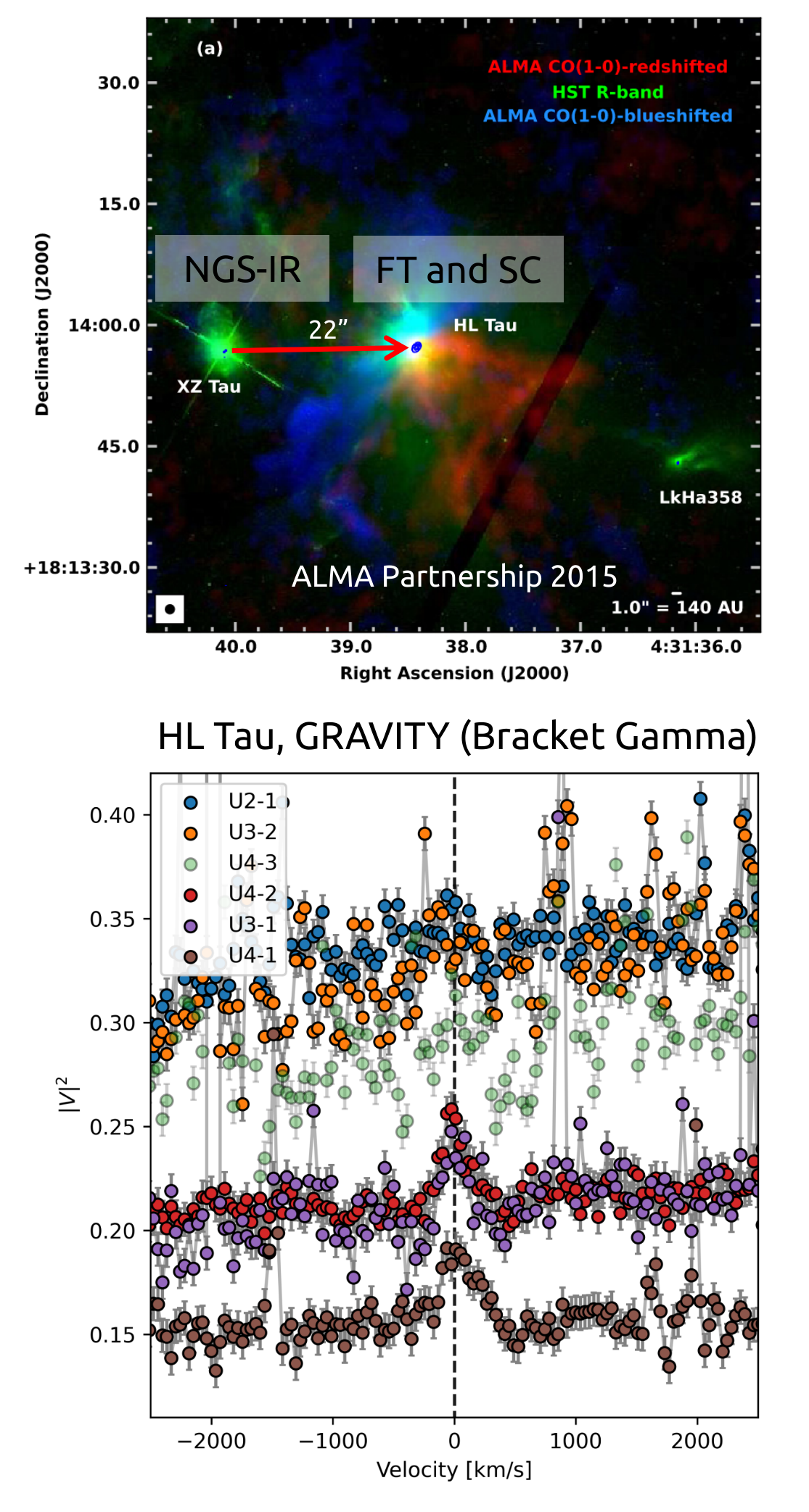}
    \caption{\textbf{Upper panel}: Overview of the HL Tau region with ALMA and HST, adapted from \cite{ALMApartnership_2015}. The NGS IR WFS was used on XZ\,Tau located at 22\arcsec{} for the AO loop, while fringe-tracking and science was done on HL\,Tau with GRAVITY. \textbf{Lower panel}: Squared visibility of HL Tau, centered around the Br-$\gamma$ line at \SI{2.16}{\micro\meter}.}
    \label{fig:SCIENCE_HLTau}
\end{figure}

\subsection{Embedded young stellar objects}

The study of the inner astronomical unit of young stellar objects has advanced significantly thanks to infrared interferometry \citep{Lazareff2017,Perraut2019,Benisty2023}, but was essentially limited to Class II objects. This limitation originates from the extinction of the envelope surrounding these systems, which are bright in the K band but faint in the G band. The GPAO-LGS mode removes this limitation on the color of the AO target, making accessible Class\,I and T\,Tauri stars on statistical scale. The NGS\_IR mode enables these observations as well, in the specific case where a bright star in the K-band is located in the vicinity of the science target ($\sim20\arcsec{}$). 

We used GPAO in the NGS\_IR and GRAVITY to observe the protoplanetary disk in \object{HL Tau}. HL\,Tau is a T\,Tauri star ($M= 1.7\,\Msun$), located in Taurus ($d=\SI{147}{pc}$). This system has been a prime target of sub-millimetric interferometry and ALMA \citep{ALMApartnership_2015}, but was never observed with VLTI due to its extremely red color ($\Rmag-\Kmag=7.0$). The GRAVITY observations were performed by closing the NGS IR WFS loop on the nearby star XZ\,Tau ($\mathrm{m}_H = 8.15 $), located at a separation of 22.5\arcsec{}. HL Tau was observed on 14 December 2024, with \SI{30}{\minute} integration on source and a typical seeing $0.5-0.7\arcsec{}$. The observations were carried out at a high-spectral resolution $R\sim 4000$, which also allows spectro-astrometry of Br-$\gamma  $ to be obtained. 

\begin{figure}[t]
\centering
\includegraphics[width=\columnwidth]{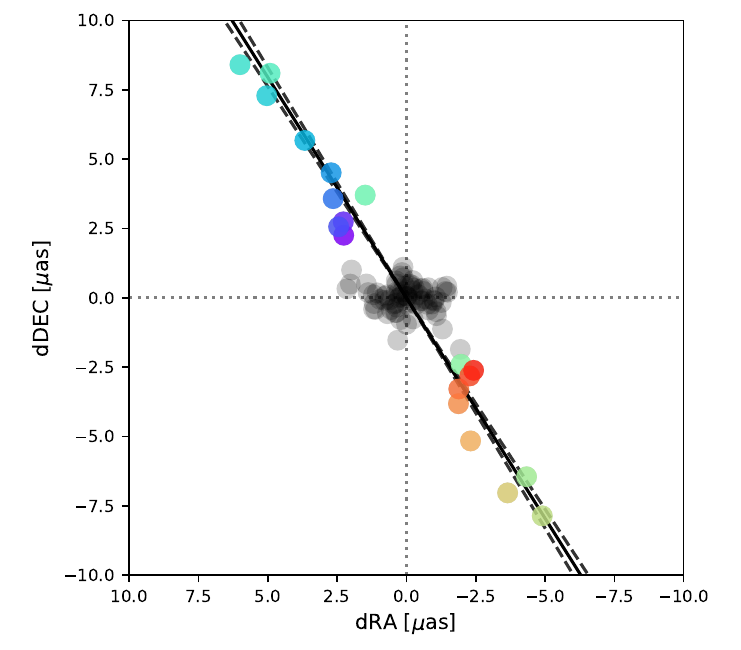}
\caption{Spectro-astrometry across the photospheric Br-$\gamma$ absorption line of the star $\beta$~Pictoris. Blue to red colours represent the velocity channels from -400 to +\SI{400}{\kilo\meter\per\second}. The gray points are the velocity channels in the continuum. The solid line is the best fit position angle for the rotation, with its uncertainty in dashed lines.}
\label{fig:SCIENCE_betaPic}
\end{figure}

In the continuum, the visibility amplitudes show that the inner disk of HL Tau is spatially resolved with GRAVITY, with visibility amplitudes smaller than 1 and consistently decreasing for the longer baselines U1-U4, U2-U3. The preliminary fit of the GRAVITY data converges to a typical half-radius of $a\sim\SI{2.0}{\milli as}$, which is consistent with the values extrapolated from T\,Tauri stars \citep{Gravity2021_ysottau}. In addition, the differential amplitudes in Br-$\gamma$ (Fig. \ref{fig:SCIENCE_HLTau}, lower panel) show that the magnetospheric region in this star is resolved. The differential spectro-astrometry provides a typical precision of $<1\deg$ in HL\,Tau, which makes it possible to probe the kinematics of Br-$\gamma$ in the magnetospheric region.
The visibility amplitude also includes a significant contribution of extended flux as probed on the short baseline U1-U2 and U3-U2. This halo corresponds to an envelope in Class\,I objects, which can be seen for this object on the GRAVITY Acquisition Camera (not shown here) and on near-infrared and ALMA images (Fig. \ref{fig:SCIENCE_HLTau}, upper panel). These observations are the first resolved observations of the inner disk of HL\,Tau below the astronomical scale unit and the first resolved observations of its magnetospheric accretion region.
The detailed analysis of these data is beyond the scope of this paper and will be presented in an upcoming publication. 

These observations demonstrate the feasibility of studying T\,Tauri stars and Class\,I objects with GRAVITY+. This will be greatly generalized with GPAO-LGS, which will enable virtually the whole population of T\,Tauri stars to be covered. 

\subsection{Sub-micro-arcsecond spectro-astrometry}

We illustrate the performance on bright targets with data recorded on the star $\beta$~Pictoris (\object{HD 39060}). This star was used as a reference during an observation of the planet $\beta$~Pic~b, for a total integration time on the star of \SI{6}{\minute}. We analyzed the data following the same methodology as \citet{LeBouquin2009}, later applied on $\beta$~Pictoris \citep{2020ApJ...897L...8K}. Figure~\ref{fig:SCIENCE_betaPic} shows the resulting spectro-astrometric signal across the photospheric Br-$\gamma$ absorption line. The rotation of the photosphere is very well resolved, with a spectro-astrometric precision of \SI{0.6}{\micro as} (1-$\sigma$) per spectral channel at $R\approx4000$. From our record, this is the first measurement with sub-micro arcsecond precision in the optical. Thanks to this exquisite precision, the position angle of the stellar rotation axis is pinned down to $32.09\pm0.9$\,deg (east of north). It allows us to answer a long standing question in this system: the stellar rotation is better aligned with the inner warped secondary disk close to the star, oriented along $33.3\pm0.6$\,deg, rather than with the extended primary disk, oriented along $29.3\pm0.3$\,deg \citep{2012A&A...542A..40L}.

\section{Conclusion and prospects}

This paper presents the design and implementation of the GRAVITY Plus Adaptive Optics for the VLTI, which introduces  an extreme AO (NGS mode) and a faint AO (LGS mode) capability to optical interferometry for the first time. This paper demonstrates the performance obtained with extreme AO, which achieves fringe-tracking up to magnitude $\Kmag=12.5$, wide-field fringe-tracking, spectroscopy on $\Kmag\approx 21$ substellar objects with $R=500$, an improvement of a factor of 10 in sensitivity on exoplanet observations, sub-micro arcsecond spectro-astrometry, and fringe-tracking residuals $\approx \SI{100}{\nano\meter}$ on the UTs for bright targets. 

GPAO-NGS will be the workhorse of the VLTI high-contrast mode for exoplanet observations in the years to come. In terms of future research, the first avenue is the implementation of wavefront control and so-called dark-hole techniques \citep{2022SPIE12183E..0VP,Pourre2024}. This method consists in injecting modal offsets into the AO to null the starlight at the exact location of the SC fiber. The improvement will benefit particularly short-separation planets with $10^{-6}$ contrast at $\approx \SI{100}{\milli as}$. This range of separation is particularly synergetic with the discovery space of Gaia DR4. Even further, the GPAO-NGS will be essential for possible upgrades of VLTI to shorter wavelengths in the J band, by upgrading the system to \SI{1.5}{\kilo\hertz} and 900 corrected KL modes. Additional prospects include the recommissioning of secondary guiding between the Coud\'e focus and the interferometric laboratory \citep{Pfuhl2014} to correct for the VLTI tunnel turbulence.

The deployment of the laser guide stars on all UTs in at the end of 2025 \citep{Messenger2022, 2024SPIE13096E..9AB} will provide the complete implementation of GPAO. The GPAO-LGS mode will make it possible to perform AO correction up to magnitude $\Grp\approx17$ with a visible tip-tilt star or up to $\Kmag\approx12$ with an infrared tip-tilt star. Combined with the wide-field fringe-tracking capability, the LGS mode will immediately increase the sky coverage available at VLTI by two orders of magnitude. For future research, the design of the low-order NGS wavefront sensors of GPAO-LGS enables many more modes than tip, tilt, and focus to be controlled (up to 12 modes in LO VIS and up to 44 in LO IR). The pros (reduced cone effect) and cons (added measurement noise) of increasing this number of modes will be explored. This is especially promising for the characterization of exoplanets around embedded young stars.

The combined GRAVITY+ upgrades bring VLTI UTs to a limiting magnitude of $\Kmag\approx 21$ (1h integration time in medium spectral resolution, under optimal conditions). This is getting close to the fundamental background limit that stands at $\Kmag\approx 25$ (see Appendix~\ref{appendix:fundamental_limits}). The worst offender is the transmission, which is  shared equally between the infrastructure and the instrument. The tracking limit at $\Kmag\approx 12.5$ still falls far short from its fundamental quantum limit of $\Kmag\approx 20$, with a supplementary offender being the vibrations of the infrastructure.

To conclude, the upgrade of VLTI with extreme AO and the upcoming deployment of the LGS mode is a true paradigm shift for observations of the optical Universe at very high angular resolution. It opens up a new parameter space for the spatially resolved study of the Galactic Center, extragalactic science at high and low redshift, exoplanets at high-contrast, the inner region of protoplanetary disks, and more cases that are yet to be unveiled. With the synergy between LGS-assisted, high performance AO and interferometry, the Paranal Very Large Telescope is reaching the most advanced vision of its initial founders (see the \emph{VLT Blue Book}\footnote{https://www.eso.org/public/products/books/book\_0005}, 1987). The time has come to imagine what could be the next vision for spatially resolved, high sensitivity, and high contrast optical astronomy.

\begin{acknowledgements}
We are very grateful to our funding agencies (MPG, ERC, CNRS [PNCG, PNGRAM], ANR [contracts: AGN Melba ANR-21-CE31-0011, ExoVLTI ANR-21-CE31-0017], European Union’s Horizon 2020 research and innovation programme [grant agreements: ORP RadioNet Pilot - No 101004719, SCIFY - No 866070, UniverScale - No 951549], DFG, BMBF, Paris Observatory [CS, PhyFOG], Observatoire des Sciences de l’Univers de Grenoble, Universite Grenoble Alpes, Observatoire de la Côte d'Azur, Universit\'e C\^ote d'Azur, R\'egion Sud, Fundação para a Ciência e Tecnologia, the Science Foundation Ireland [Grant No. 18/SIRG/5597]), the Research Foundation - Flanders (FWO) [Grant No. 1234224N], R\'egion Auvergne-Rhone-Alpes, and the generous support from the Max Planck Foundation - an independent, non profit organization of private supporters of top research in the Max Planck Society. JBLB thanks E.~Gendron and Y.~Clenet. F.M would like to thank L. Perez. The community is encouraged to prepare their future VLTI observations with the dedicated tools developed by the \texttt{JMMC}, in particular \texttt{searchfft}\footnote{https://searchftt.jmmc.fr}, \texttt{aspro2}\footnote{https://www.jmmc.fr/aspro} and \texttt{searchCal}\footnote{https://www.jmmc.fr/searchcal}, which were also used in this work. This research made use of the SIMBAD database, operated at CDS, Strasbourg, France. This work made use of \texttt{Astropy}\footnote{https://www.astropy.org/} a community-developed core Python package and an ecosystem of tools and resources for astronomy (Astropy Collaboration et al. 2013, 2018, 2022). We thank the anonymous referee for his comments, which helped improve this article. We express our sincere gratitude to the operational staff at ESO/Paranal and ESO, as well as the scientific, administrative, and technical staff at our institutions, who have made GPAO a reality.
\end{acknowledgements}

\bibliographystyle{aa}
\bibliography{aa55666-25}

\begin{appendix}

\section{Conceptual differences with ERIS}
\label{appendix:ERIS}

Although GPAO shares a large number of requirements and constraints with ERIS and AOF, there is a set of differences with significant impact on the design choices and development strategy. (I)~The DM of GPAO does not have an absolute feedback position sensor. The actual DM shape can only be inferred from the command vector sent to the DM. (II)~In GPAO, the DM (located in M8) and the photometric pupil (located in M2) are two different optical surfaces, linked with a variable wobble and rotation angle (the elevation). (III)~GPAO has access to the Nasmyth beacon of the Unit Telescope, located before the DM (near M4). It allows closed loop tests and calibration of the AO during daytime. ERIS and the AOF instruments do not have access to this tool: AO calibrations and tests can only be executed at night, on star light. (IV)~GPAO only has a loose requirement regarding the absolute position and the slow drift of the field sent to the science instrument ($<0.2\arcsec{}$ for GPAO with respect to $<0.01\arcsec{}$ for ERIS). Indeed, the VLTI provides a secondary tip-tilt guiding to GPAO, at a rate of \SI{1}{\hertz}, ensuring a constant fine alignment of the field. (V)~Interferometric fringe tracking requires a diffraction-limited beam, at least partially. The disappearance of a coherent core in the delivered PSF corresponds to the practical limiting magnitude. Therefore, there is no such modes as ``seeing enhancer'' (only LGS correction) or ``seeing limited'' (no AO correction) in GPAO.

\section{Parameters of the PSIM model}
\label{sec:PSIM_parameters}

\begin{table*}[h]
\centering
\caption{Parameters of the PSIM model for the four GPAOs.}
\label{tab:PSIM_parameters} 
\begin{tabular}{l c c c c c c}
\hline\hline
  \text{Parameter}  &  \text{Description} & \text{Unit} & \text{GPAO1} & \text{GPAO2} & \text{GPAO3} & \text{GPAO4}  \\
\hline
COMMON\\
DM inclination & design &  deg & 13.26 & 13.26 & 13.26 & 13.26 \\
DM stroke\tablefootmark{a} & measured & \SI{}{\micro\meter} & 3.9 & 3.7 & 3.8 & 3.9 \\
\hline
NGS VIS (measured in HO $40\times40$)\\
DM/WFS pitch scaling & measured & \% & 102.8 & 103.5 & 102.1 & 103.5 \\
DM/WFS registration angle & $az$-12.984+ & deg & 0.60 & 1.21 & -1.14 & -2.95 \\
DM/WFS anamorphosis\tablefootmark{b} 0 & measured & \% & 0 & 0 & 0 & 0 \\
DM/WFS anamorphosis\tablefootmark{b} 45 & measured & \% & 0 & 0 & 0 & 0 \\
Photometric pupil scale & measured & \% & 97.0 & 96.9 & 96.8 & 97.1 \\
\hline
LGS\\
DM/WFS pitch scaling & measured & \% & 101.3 & 102.2 & 101.2 & 102.0 \\
DM/WFS registration angle & $az$-12.984+ & deg & -88.00 & -91.74 & -88.65 & -88.35 \\
DM/WFS anamorphosis 0 & measured & \% & -0.2 & -0.1 & -0.2 & -0.2 \\
DM/WFS anamorphosis 45 & measured & \% & 0.1 & 0.0 & 0.1 & 0.0 \\
Photometric pupil scale & measured & \% & -- & -- & -- & 99.0 \\
\hline
NGS IR\\
DM/WFS pitch scaling & measured & \% & 96.4 & 95.4 & 101.1 & 97.7 \\
DM/WFS registration angle & $az$-12.984+ & deg & 78.71 & 80.01 & 83.81 & 81.11 \\
DM/WFS anamorphosis 0 & measured & \% & 0.6 & 3.7 & -0.4 & 0.3 \\
DM/WFS anamorphosis 45 & measured & \% & 0.6 & -0.6 & -0.8 & 0.0 \\
\hline
\end{tabular}
\tablefoot{\tablefoottext{a}{In PSIM, the DM stroke is defined as the maximum mechanical deformation of the surface when pushing a single actuator, with respect to the reference. This mechanical stroke is achievable toward the positive and toward the negative direction.} \tablefoottext{b}{The anamorphoses of the NGS VIS wavefront sensor were measured to be negligible with a fixed offset and an azimuthal contributions below 0.04 \% each. We consequently let them equal to 0\% in the PSIM model.}}
\end{table*}

Table~\ref{tab:PSIM_parameters} sums up the parameters of the PSIM model of the different GPAOs, namely (i) the DM amplitude, and (ii) the DM/WFS mis-registration parameters: the DM pattern magnification and anamorphoses (along the 0 deg and 45 deg axes) on the WFS and the offset angle between the DM and the WFS (the actual angle between the DM and WFS evolving with the azimuth $az$). All these parameters were fitted on the UT Nasmyth beacon. The on-sky photometric pupil scale is not a PSIM parameter but is indicated for information. For the LGS WFS, it is yet to be measured on-sky for GPAO1-2-3. The magnification of the photometric pupil in the NGS IR WFS is not fitted due to the lack of sensitivity of the $9 \times 9$ SH-WFS.

\section{Secondary loops}

\label{app:SL}
The Secondary Loops (SL) run in the OS software, typically at a period of a few seconds. The function of most of them has already been introduced in Sect.~\ref{sec:control_strategy}. We provide here a simple but exhaustive summary.

\paragraph{SL1} offloads the DM shape in the active optics of the M1 mirror of the UT. This loop has been decommissioned because the DM has enough stroke.
\paragraph{SL2} guides the DM registration on the NGS VIS WFS using the NPSM.
\paragraph{SL3} guides the DM registration on the LGS WFS, using the LPSM.
\paragraph{SL4} calculates the NCPA modal vector and reference slopes to be sent to the SRTC. The NCPA vector contains a static part and, when observing off-axis, a dynamical part related to the aberrations across the M9 DIC and PBS.
\paragraph{SL5} updates the rotation angles of the PSIM model in the SRTC, which triggers the recomputation of the Control Matrices. It also updates the rotation angles of various monitoring components.
\paragraph{SL6} guides the DM registration on the NGS IR WFS using the Variable Curvature Mirror (VCM) of the STS.
\paragraph{SL7} offloads the focus measured in the LGS WFS into the LFOC device.
\paragraph{SL8} offloads the mean tip and tilt components of the DM shape. The offload goes to the telescope axes when observing on-sky or goes to the QSM when using the Nasmyth beacon.
\paragraph{SL9} implements the pointing model and the tracking trajectory for the laser Jitter Mirror, by regularly updating the SRTC. Doing so, it ensures that the laser spot remains inside the LGS WFS even when the Jitter loop is open.

\section{Data flow of the HRTC}
\label{appendix:saturation}

\begin{figure*}[ht]
\centering
\includegraphics[width=\textwidth]{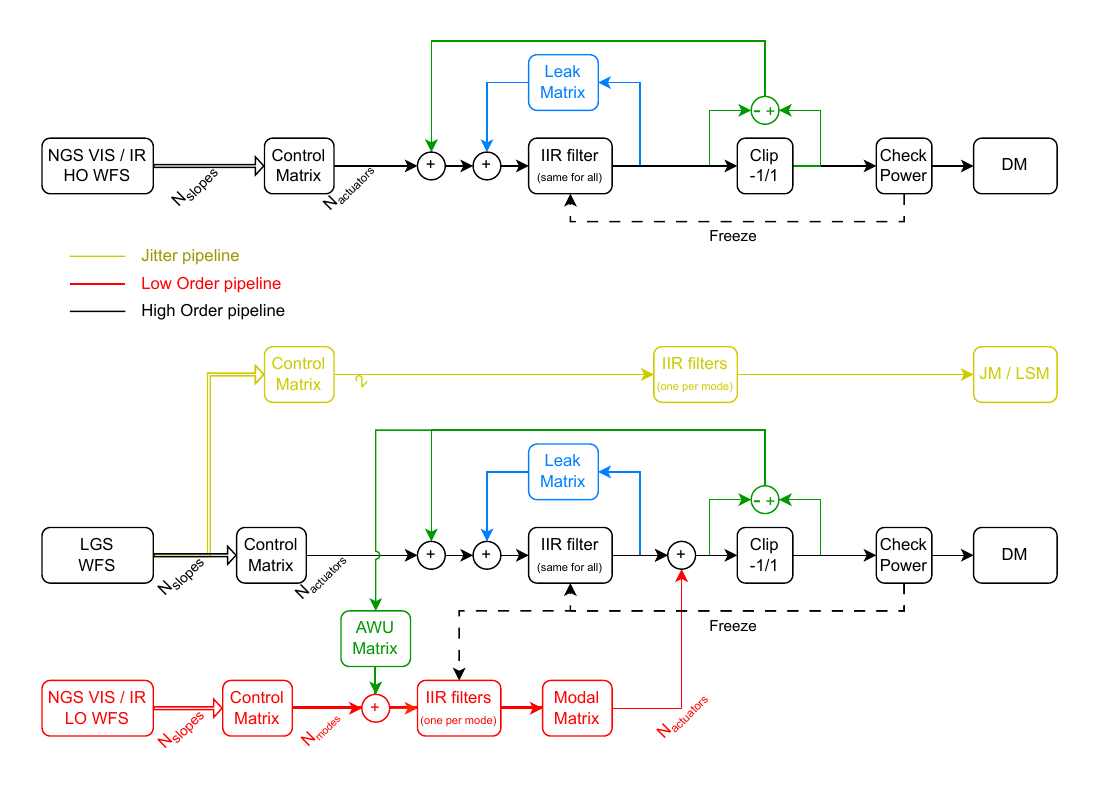}
\caption{Simplified diagram of the real-time data-flow within the HRTC of GPAO when configured in NGS\_VIS/IR mode (top) and in LGS\_VIS/IR mode (bottom), from the slopes vectors computed by the Wavefront Sensors to the command vectors sent to the Deformable Mirror and to the Jitter Mirror. The black path is the HO pipeline, the red path is the LO pipeline, the yellow path is the Jitter pipeline. Each pipeline can be closed independently. The green feedback is the anti-windup and the blue feedback is the modal leak (which also acts as a garbage collector). To ease the reading, the diagram does not include the pre-processing from detector images to slopes, it does not include the reference slopes offsets, and it does not include the real-time disturbances (in slopes and command spaces) than can be input to the data-flow for calibration purposes.}
\label{fig:saturation}
\end{figure*}

Figure~\ref{fig:saturation} presents a simplified diagram of the data flow in the GPAO HRTC. The high-order controllers are implemented as "implicitly modal": the input slope vector is converted into a delta-command vector in DM space by the matrix multiplication with the Control Matrix. These delta-commands are filtered by a temporal IIR filter, identical for all elements of the delta-command vector (as-of-now, the IIR is configured as a pure integrator). On the other hand, the low order controllers are implemented as "explicitly modal": the temporal IIR filter acts on the delta-command expressed in KL modes. The IIR filter is possibly different for each mode. A second matrix multiplication then projects the modes into the DM actuator space.

If needed, the commands from the low order and the high-order pipelines are summed. The output command vector is clipped to +1/-1, in unit of DM maximum stroke. The total power of the command ($\sum c_i^2$) is compared to the maximum power allowed by the DM electronic, and if larger, the command is not send to the DM and the IIR filter is frozen. Clipping (e.g. zonal saturation) regularly occurs on a few actuators on the edges when the seeing goes above 1.5\arcsec{}. Freezing (e.g., power saturation) only occurs in pathological cases such as a wrong alignment. Note that clipping a few actuators has a negligible impact on the delivered image quality, but freezing the command is generally dramatic.

GPAO implements the classical anti-windup method: the clipped part of the command vector (if any) is subtracted to the delta-command of the next loop-cycle. It prevents from large accumulation of uncorrected commands inside the controller ("windup"), which would take a long time to reset when the disturbance actually comes back inside the dynamic range. Simulations demonstrated the importance of this anti-windup path.

GPAO also implements an innovative modal leak path: the command actually sent to the DM is filtered through a leak matrix and subtracted to the delta-command of the next loop-cycle. This leak matrix is constructed in a way to (1) fully leak the uncontrolled space and (2) partially leak the high-order controlled modes. Mathematically, let's call $A2S$ the interaction matrix (actuators to slopes) and $M2A$ the modal matrix (modes to actuators). Let's call $G_m$ a vector of modal gain, and $L_m$ a vector of modal leak (typically 0 for the first modes and 0.5 for the highest controlled modes). Then, the high-order control matrix writes
 \begin{equation}
   CM = M2A \times G_m \times (A2S \times M2A)^{-1} \;,
 \end{equation}
 and the leak matrix writes
 \begin{equation}
   LM = \identity - M2A \times (\identity - L_m) \times M2A^{-1} \;.
 \end{equation}
where $\identity$ is the identity matrix. The setup of the modal leak vector $L_m$ in GPAO follows the prescriptions of \citet{2019arXiv191105989A}. Note that the size of the high-order control matrix is  $(N_{slopes}\times N_{actuators})$ while the size of the leak matrix is $(N_{actuators}\times N_{actuators})$. The computations of the anti-windup and of the leak paths are executed in the dead-time of the HRTC, that is once all computations depending on the pixels of the current frame are finished and before the pixels of the next frame arrive. Thus, these computations do not contribute to the overall latency. 

The anti-windup does not contain any projection, and thus creates signal outside the control space, often called ``garbage''. Nevertheless, there is no need for a dedicated garbage collector as in SPHERE \citep{2014SPIE.9148E..1UF} because the uncontrolled space is quickly erased by the $LM$ projection at frame rate.

\section{Transmission}
\label{appendix:transmission}

\begin{figure}[ht]
\centering
\includegraphics[width=\columnwidth]{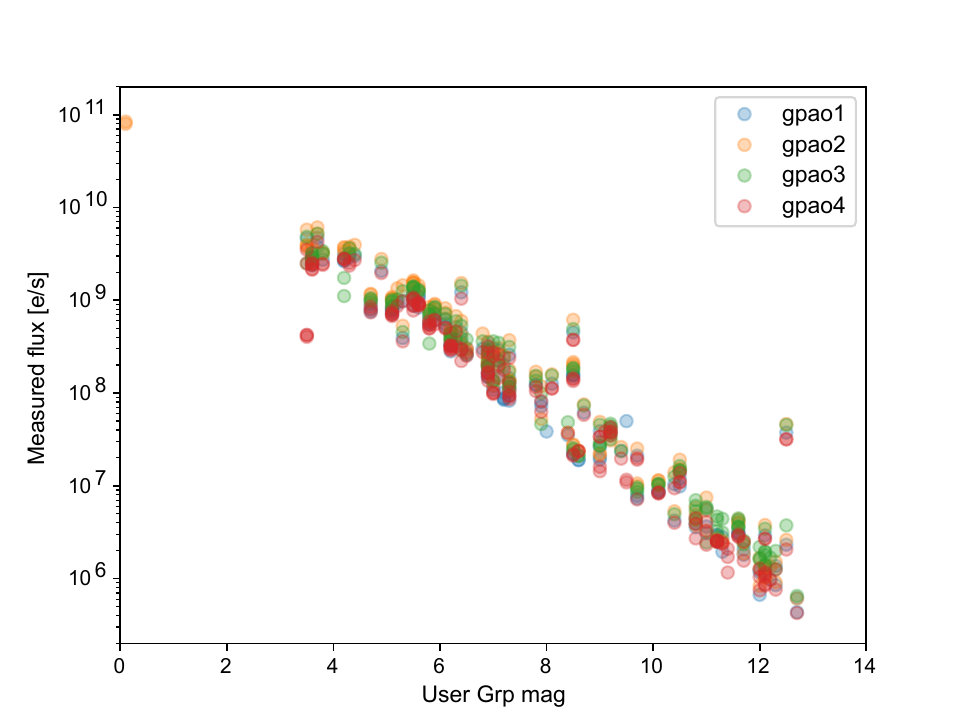}
\caption{Measured total flux on the visible WFS of GPAO versus the \Grp{} Gaia magnitude.}
\label{fig:GPAO_NGS_VIS_flux_vs_mag}
\end{figure}

\begin{figure}[ht]
\centering
\includegraphics[width=\columnwidth]{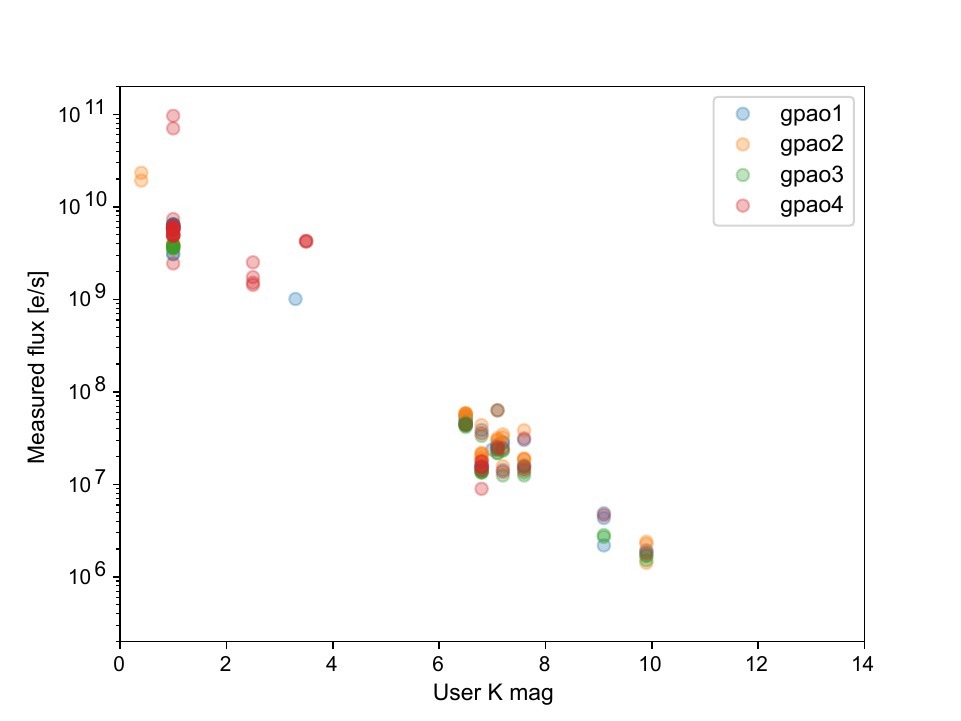}
\caption{Measured total flux on the infrared NGS IR WFS versus the K band magnitude.}
\label{fig:GPAO_NGS_IR_flux_vs_mag}
\end{figure}

Figure~\ref{fig:GPAO_NGS_VIS_flux_vs_mag} and Figure~\ref{fig:GPAO_NGS_IR_flux_vs_mag} show the total number of electrons received on the detector of the visible WFS and of the infrared WFS as a function of the Gaia red-pass (\Grp{}) or K-band magnitude of the guide star. The figures include all science observations from December 2024 to February 2025. We consider a detector conversion gain of 30\,e/adu for the OCAM2 camera and 10\,e/adu for the Saphira IR detector. The measurements are corrected for the amplification gain and framerate.

\section{Transfer function}
\label{app:transfer_function}

\begin{figure}[ht]
\centering
\includegraphics[width=\columnwidth]{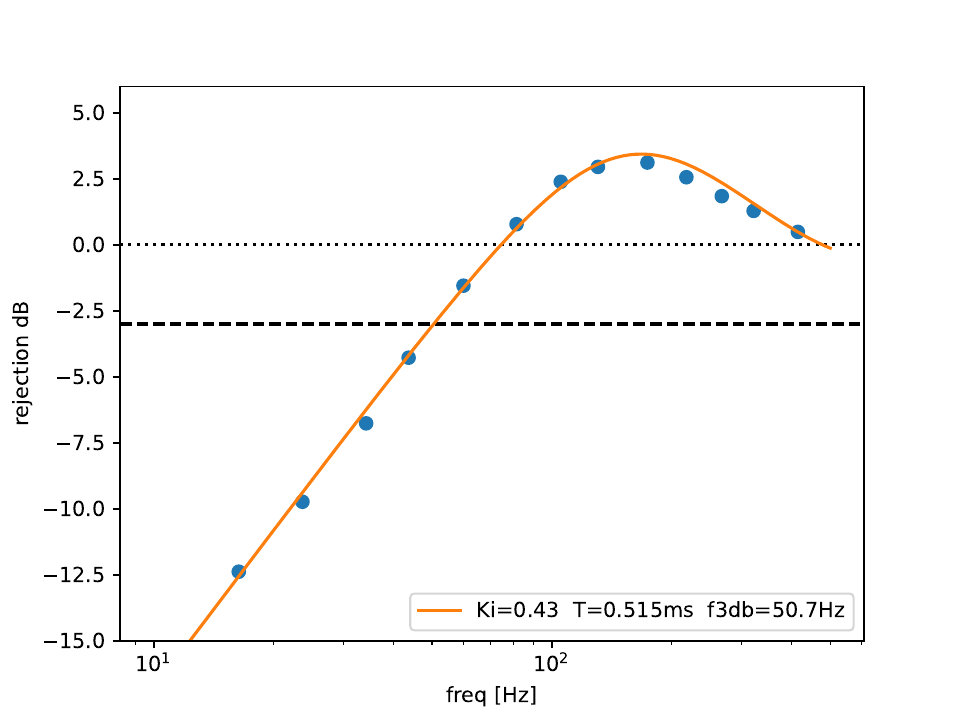}
\includegraphics[width=\columnwidth]{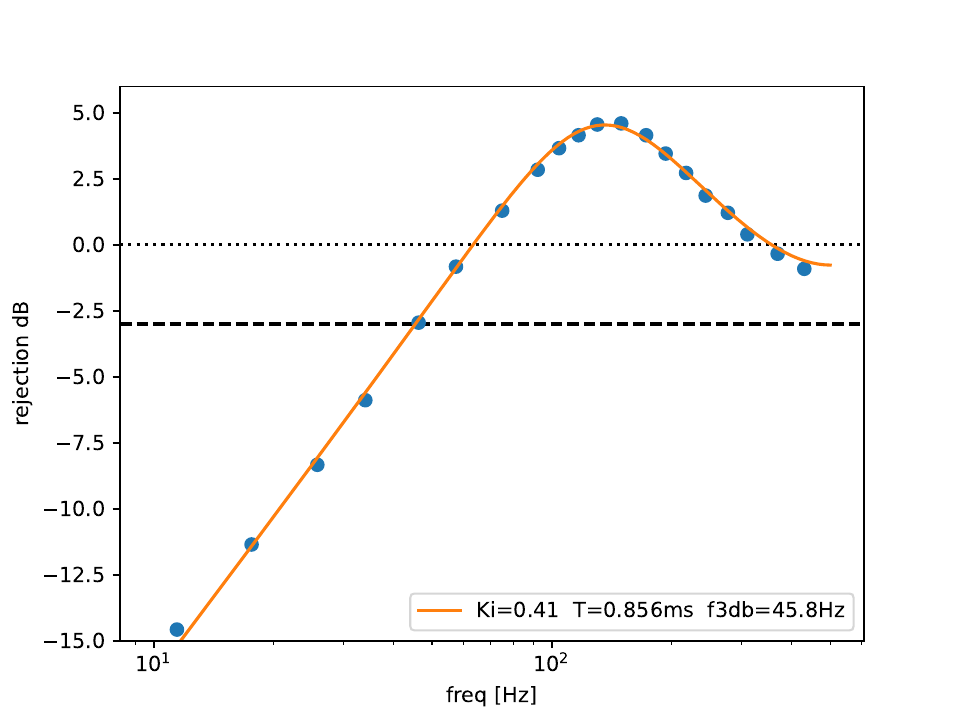}
\caption{Measurements (dots) and models (lines) for the wavefront transfer function for the infrared (top) and visible WFS (bottom), configured at the maximum frame rate of \SI{1000}{\hertz}. The model is fitted to the data by adjusting the integral gain $K_i$ and the pure-delay $T_d$. The rejection bandwidth defined at -3\,dB is also given. Data are presented for GPAO2 but similar results are obtained on the other systems.}
\label{fig:GPAO_TF}
\end{figure}

The DM is modeled with a sum of 3 second-order resonances with frequencies $s_i=\{1055, 1265, 1600\}$ \SI{}{\hertz}, damping coefficients $\zeta_i = \{0.025, 0.025, 0.1\}$ and amplitudes $a_i=\{0.33, 0.4, 0.27\}$:
\begin{equation}
DM(s) = \sum_i \frac{a_i}{(\frac{s}{s_i})^2+2\zeta_i\frac{s}{s_i}+1}
\end{equation}

The WFS is modeled as a discrete integration window of integration time $T_i$:
\begin{equation}
WFS(s) = \frac{1-e^{-s\,T_i}}{s\,T_i}
\end{equation}

The controller is modeled as a discrete integrator with gain $K_I$ and frame period $T_c$:
\begin{equation}
C(s) = \frac{K_I}{1-e^{-s\,T_c}}
\end{equation}

The output command is hold for one frame period $T_c$ resulting in an additional:
\begin{equation}
H(s) = \frac{1-e^{-s\,T_c}}{s\,T_c}
\end{equation}

All other delays are factorized into a pure-delay $T_d$:
\begin{equation}
D(s) = e^{-s\,T_d}
\end{equation}

The closed-loop rejection function writes:
\begin{equation}
R(s) = 1/(1+DM(s).WFS(s).C(s).H(s).D(s)).
\end{equation}
Measurements (dots) and models (lines) for the wavefront transfer function are presented in Fig.~\ref{fig:GPAO_TF}. Leaks were deactivated during these measurements.

\section{Fundamental limits}
\label{appendix:fundamental_limits}

This appendix recall the fundamental limits for correcting the atmosphere and integrating on the science. More details can be found in \citet{System2023}. By "fundamental" we call limits that cannot be lifted by technological improvements (atmosphere characteristics, quantum noise, integration time). We restrict this example to the K-band for concision. The photon flux in [$\gamma\,\mathrm{m}^{-2}\,\mathrm{s}^{-1}$] is estimated with the relation $n_\gamma=10^{(25.0-\Grp{})/2.5}$ in the visible and $n_\gamma=10^{(23.4-\Kmag{})/2.5}$ in the infrared. We define a S/N of 5 as threshold for the limiting magnitude, so about 25 photons when quantum noise limited. We consider a turbulence with a Fried parameter $r_0=\SI{20}{\centi\meter}$ and a coherence time of $\tau_0=\SI{10}{\milli\second}$ in the visible, and a longer coherence time of $\tau_0=\SI{40}{\milli\second}$ in the infrared. We consider a scientific integration time $T=\SI{1}{\hour}$. The infrared sky background is $\Kmag\approx21$ for a solid angle of $40\,\mathrm{mas}^2$ (acceptance of the GRAVITY single mode fiber, fixed by the \SI{8}{\meter} diameter of the UTs).

\paragraph{NGS AO limit}
The NGS adaptive optics is limited by the S/N within the coherence volume defined by $r_0^2\times\tau_0$. Getting 25 photons per volume translates into a magnitude $\Grp=13.1$.

\paragraph{LGS-assisted AO limit}
The LGS-assisted adaptive optics is limited by the S/N within the coherence volume defined by $D^2\times\tau_0$, where D is the diameter of the telescope. Getting 25 photons per volume translates into a magnitude $\Grp=20.3$ for a \SI{8}{\meter} aperture.

\paragraph{Fringe-tracking limit}
The fringe-tracking is limited by the S/N within the coherence volume defined by $\frac{2}{3}D^2\times\tau_0$ (the fringe receives the flux of 2 telescopes, each of them being split in 3 to feed all pairs). Getting 25 photons per volume translates into a magnitude $\Kmag=20.2$ for an array of four \SI{8}{\meter} apertures.

\paragraph{Science limit}
At its best sensitivity in the lowest spectral resolution, the science is limited by the S/N within the coherence volume defined by $\frac{2}{3}D^2\times T$. This is limited by the sky-background and not the quantum noise. Getting a number of photons equal to 5 times the background photon noise translates into a magnitude $\Kmag=25.2$ for an array of \SI{8}{\meter} apertures.

\paragraph{Caveats}
This analysis has two caveats. First, the spatial filtering effect across the telescope aperture increases the coherence time of the piston. Secondly, the dilution factor $\frac{2}{3}$ is partially lifted by the global retrieval of all pistons with all measurements.  As such, the above computed fringe-tracking limit is underestimated.

\section{Nomenclature}
\label{app:nomenclature}

\begin{enumerate}[leftmargin=0cm]
    \item[2D] Two dimensional
    \item[ADC] Atmospheric Dispersion Compensator, made of 2 rotating prisms.
    \item[AGN] Active Galactic Nucleus
    \item[AO] Adaptive Optics
    \item[AOF] Adaptive Optics Facility
    \item[BLR] Broad Line Region
    \item[CCD] Charge Couple Device
    \item[CIAO] Coud\'e Infrared Adaptive Optics
    \item[CM] Control Matrix
    \item[CO] Corrective Optics assembly, hosting the QSM and the DM, located in the M8 of the UT
    \item[Coud\'e] Train of mirrors M4 to M8 of the UT, co-rotating with azimuth
    \item[DCS] Detector Control Software
    \item[DM] Deformable Mirror
    \item[DCR] Dichroic splitting the blue (reflected toward the technical camera) and the red (transmitted toward the NGS WFS camera).
    \item[EMCCD] Electron-Multiplication Charge Coupled Device is a technology of fast and low noise visible camera
    \item[ERIS] Enhanced Resolution Imager and Spectrograph
    \item[ESO] European Southern Observatory
    \item[FT] Fringe Tracker
    \item[FSM] Field Steering Mirror of the STS, actuates the tip-tilt
    \item[GALACSI]  Ground Atmospheric Layer Adaptive Corrector for Spectroscopic Imaging
    \item[GbE] Gigabit Ethernet
    \item[GPAO] GRAVITY Plus Adaptive Optics
    \item[GRAAL] GRound layer Adaptive optics Assisted by Lasers 
    \item[G-Wide] GRAVITY-Wide is the new implementation of the dual beam, off-axis capability of VLTI allowing to feed the GRAVITY instrument with 2 fields separated by up to \SI{1}{arcmin}.
    \item[HO] High-order, in NGS modes and in opposition to the low-order in LGS modes.
    \item[HSDL] High Speed Data Link
    \item[HKL] House Keeping Link
    \item[HRTC] Hard Real Time Computer, performs the real-time tasks    
    \item[ICS] Instrument Control Software
    \item[IF] Influence Functions  
    \item[IIR] Infinite Impulse Response is the default type of temporal filter used in SPARTA (alternative is PID).
    \item[IM] Interaction Matrix
    \item[IR] InfraRed
    \item[IRIS] guiding camera of the VLTI
    \item[ISS] The Interferometric Supervisor Software is the high-level layer coordinating the all the subsystems of VLTI.
    \item[JM] Jitter Mirror of the LGS Launch telescope, fast actuator to stabilize the observed position of the LGS spot in the WFS sensor.    
    \item[JWST] James Webb Space Telescope
    \item[KL] Karhunen-Lo\`ve modes, in GPAO these modes are defined in DM command space.   
    \item[LCU] Local Controller Unit, local controller to interface with the hardware, generally based on VXworks in the ESO context
    \item[LFOC] Laser Focus Control, long range longitudinal motion of the entire LGS WFS assembly to maintain the focalisation toward the Sodium layer. The range extend toward infinity to allow observing the Nasmyth beacon of the UT
    \item[LROT] LGS derotator, fixed in operation
    \item[LGS] Laser Guide Star
    \item[LO] Low order, in opposition to the high order in LGS modes
    \item[LPS] Laser Projection Subunit
    \item[LPSM] Laser Pupil Steering Mirror, located in a focal plan within the LGS WFS.
    \item[LSM] Laser Steering Mirror (called FSM in ERIS/AOF), slow offload actuator for the JM. 
    \item[M1] primary mirror of the UT, slowly deformable (active optics).
    \item[M2] secondary mirror, pupil stop of the UT.
    \item[M9 DIC] Dichroic at the Coud\'e focus of the UT, splitting the IR (reflected toward STS) and the VIS (transmitted toward GPAO WFS). 
    \item[MACAO] Multi-Application Curvature Adaptive Optics
    \item[MATISSE] Multi AperTure mid-Infrared SpectroScopic Experiment
    \item[NAOMI] New Adaptive Optics Module for Interferometry, installed on the \SI{1.8}{\meter} Auxiliary Telescopes of the VLTI.
    \item[NCPA] Non common path aberrations, differential optical aberrations between the AO path and the scientific path.
    \item[NDIA] NGS diaphragm, an adjustable field stop installed in the NGS VIS WFS.
    \item[NFB] Nasmyth Focus Beacon, a set of source at the folded Nasmyth focus of the UT, feeding toward the Coud\'e train.    
    \item[NGS] Natural Guide Star 
    \item[NGS IR WFS] Infrared wavefront sensor on a natural guide star, the hardware is identical to the former CIAO WFS.
    \item[NGS VIS WFS] Visible wavefront sensor on a natural guide star. In the GPAO NGS\_VIS mode, this corresponds to the $40\times40$ SH. In the GPAO LGS\_VIS mode, this corresponds to the $4\times4$ SH.
    \item[NPSM] NGS Pupil Steering Mirror, located in a focal plan within the NGS VIS WFS
    \item[NROT] NGS derotator, fixed in operation
    \item[OPD] Optical Path Difference
    \item[OS] Observation Software    
    \item[PBS] Periscope Beam Splitter
    \item[PBS DIC] the Periscope Beam Splitter dichroic is the dichroic at the Coud\'e focus of the UT, splitting the LGS wavelength (reflected toward LGS WFS) and the broad band VIS (transmitted toward GPAO NGS VIS WFS)
    \item[PID] Proportional Integral Derivative, a type of temporal filter.
    \item[PIONIER] Precision Integrated-Optics Near-infrared Imaging ExpeRiment
    \item[PLC] Programmable Logic Controller, local controller to interface with the hardware, generally based on Beckhoff in the ESO context  
    \item[PSM] Pupil Steering Mirror, lateral pupil actuator (LPSM in the LGS WFS, NPSM in the NGS WFS).    
    \item[QSM] Quasi Static Mount, gimbal tip-tilt mount of the DM, with slow motion and large steering range.
    \item[RTC] Real Time Computer
    \item[SC] Science Channel
    \item[sFPDP] Serial Front Panel Data Port
    \item[SH] Shack-Hartmann wavefront sensor
    \item[SL] Secondary Loop, supervision, or tracking, or guiding loop executed by the AOO process of the OS. Typical rate 1 to \SI{10}{\second}.
    \item[SMBH] Super Massive Black Hole
    \item[S/N] Signal to noise ratio
    \item[SPARTA] Standard Platform for Adaptive optics Real Time Applications, developed by ESO.
    \item[SPHERE] Spectro-Polarimetic High contrast imager for Exoplanets REsearch
    \item[SRTC] Soft Real Time Computer, perform the supervision tasks
    \item[STS] Star Separator, located in the VLT Coud\'e room, split the field-of-view into two region fed to the VLTI    
    \item[UT] Unit Telescope.
    \item[VCM] Variable Curvature Mirror of the STS, also serving as lateral pupil actuator.
    \item[VIS] Visible
    \item[VLT] Very Large Telescope.
    \item[VLTI] Very Large Telescope Interferometer.
    \item[WFS] WaveFront Sensor
    \item[XYT] X,Y Table, 2D translation stage to patrol the entire field-of-view at the Coud\'e focus with the NGS WFS (NXYT) and with the LGS WFS (LXYT)
    \item[YSO] Young Stellar Object
\end{enumerate}

\end{appendix}

\end{document}